\documentclass[twocolumn, superscriptaddress,nofootinbib]{revtex4-2}

\usepackage[utf8]{inputenc}

\usepackage{amsmath, amsfonts, amssymb}
\usepackage{mathtools}
\usepackage{mathrsfs}
\usepackage{yhmath}
\usepackage{amssymb}
\usepackage{tensor}
\usepackage{accents}
\usepackage{tipa}
\usepackage{comment}
\usepackage{braket}
\usepackage[hidelinks]{hyperref}
\newcommand{\ee}{\mathrm{e}}

\usepackage{lineno}
\newcommand{\diff}{\mathrm{d}}
\newcommand{\qmarks}[1]{``#1''}

\begin{document}

\title{Cosmic Acceleration from Quantum Gravity:\\Emergent Inflation and Dynamical Dark Energy}
\author{Luca Marchetti}
\email{luca.marchetti@oist.jp}
\affiliation{
Kavli Institute for the Physics and Mathematics of the Universe (WPI),\\ UTIAS, The University of Tokyo, Chiba 277-8583, Japan}
\affiliation{Okinawa Institute of Science and Technology Graduate University,\\
Onna, Okinawa 904 0495 Japan}
\author{Tom Ladst{\"a}tter}
\affiliation{Arnold Sommerfeld Center for Theoretical Physics, Ludwig-Maximilians-Universit{\"a}t M{\"u}nchen,\\ Theresienstrasse 37, 80333 M{\"u}nchen, Germany}

\author{Daniele Oriti}
\affiliation{Departamento de Fisica Teorica \& IPARCOS, Facultad de Ciencias Fisicas,\\ Universidad Complutense de Madrid, \\ Plaza de las Ciencias 1, 28040 Madrid, Spain, EU}
\date{\today}

\begin{abstract}
We present a mechanism for the emergence of cosmic acceleration within the mean-field approximation of Group Field Theory models of quantum gravity. Depending on the interaction type, the resulting cosmological dynamics can either feature a late-time attractor corresponding to a dynamical dark energy phase—often with characteristic phantom behavior, including in models inspired by simplicial gravity—or instead support an early slow-roll inflationary epoch driven by the same underlying quantum-gravitational effects. This emergent inflation, effectively captured by a single-field description, can sustain the required expansion, naturally avoids the graceful exit problem, and appears to transition into a persistent, non-accelerating phase consistent with classical expectations.  
\end{abstract}
\maketitle
\section{Introduction}
The $\Lambda$ cold dark matter ($\Lambda$CDM) model is widely acknowledged as the standard model of cosmology. It is embedded within our best theory of classical gravity, General Relativity (GR), and it rests on the ad hoc introduction of dark energy (DE) in
the form of a cosmological constant ($\Lambda$), driving late-times acceleration, together with a cold dark matter (CDM)
component underlying large-scale structure formation \cite{Dodelson:2020bqr}. In addition, an inflationary phase of accelerated expansion at early times—often driven by scalar field, the inflaton—is invoked to address problems of standard
cosmology, while also seeding the origin of structure \cite{Starobinsky:1980te, guth,LINDE1982389,Riotto:2002yw}. Despite its remarkable phenomenological success,
$\Lambda$CDM can only be regarded as an effective model. Indeed, it faces intrinsic theoretical limitations, all suggesting a quantum gravity completion: it entails an
initial singularity where GR itself breaks down; inflation is highly sensitive to high-energy (especially trans-Planckian) physics \cite{Martin:2000xs}; theoretical estimates of the value of $\Lambda$ are not reliable \cite{Martin:2012bt}. 

Beyond these theoretical shortcomings, increasingly precise observations are exposing tensions within the $\Lambda$CDM
paradigm, even as an effective framework. A prominent example is the $H_0$ tension \cite{DiValentino:2021izs}, a $\sim 4–6\sigma$ discrepancy between determinations of the present-day
expansion rate from local standard candles and from the cosmic microwave background (CMB).
Further signs of discordance are emerging from the latest DESI results, which disfavor a pure cosmological constant at the $2.8$–$4.2\sigma$ level and instead point toward a dynamical dark-energy component, potentially with “phantom’’ characteristics \cite{DESI:2025zgx}. Motivated mostly by the $H_0$ tension and the DESI results, a wide range of phenomenological models for dynamical dark energy have been explored \cite{Avsajanishvili:2023jcl}. However, these constructions often lack a clear physical underpinning, which highlights the necessity of treating dark energy as a dynamical sector whose behavior must ultimately be derived from new physical principles. 

Analogous tensions are also appearing in early-Universe cosmology. Measurements of CMB polarization continue to yield a surprisingly low amplitude of primordial gravitational waves compared with the expectations of the inflationary paradigm \cite{Planck:2018jri}. Moreover, combined ACT–Planck analyzes are placing considerable pressure on even the better-motivated inflationary models \cite{ACT:2025tim}. Explaining these data seems to require increasingly contrived inflationary scenarios with finely tuned potentials, raising questions on the physical origin of such inflationary models.

Taken together, these theoretical challenges and persistent tensions with observations highlight the need for a more refined physical picture of the universe, especially regarding the mechanisms that drive cosmic acceleration, which lie at the heart of dark energy and inflationary physics.

Recent developments suggest a deep connection between cosmic acceleration and quantum gravity \cite{deCesare:2016rsf,Oriti:2021rvm,Pang:2025jtk} within the tensorial group field theory (TGFT) framework~\cite{Freidel:2005qe,Oriti:2006se,Oriti:2011jm,Carrozza:2013oiy,Carrozza:2016vsq,Gielen:2024sxs,Marchetti:2024tjq}. Tensorial Group Field Theories (TGFTs) extend matrix and tensor models into a fully field-theoretic setting. When the fields are endowed with quantum geometric data encoded in group-theoretic variables, the resulting models are known as Group Field Theories (GFTs). They provide a formulation of quantum and statistical field theory for spacetime quanta, where the fields live on a group manifold rather than a spacetime manifold. This quantum geometric foundation establishes close links with (and in fact is often motivated by) other non-perturbative quantum gravity approaches, including ~\cite{Ashtekar:2004eh,Oriti:2013aqa,Oriti:2014yla}, spin foam models~\cite{Perez:2003vx,Perez:2012wv}, simplicial gravity~\cite{Bonzom:2009hw,Baratin:2010wi,Baratin:2011tx,Baratin:2011hp,Finocchiaro:2018hks} or dynamical triangulations~\cite{Loll:1998aj,Ambjorn:2012jv,Jordan:2013sok,Loll:2019rdj}.

Spatially homogeneous and isotropic cosmological dynamics (with scalar field matter) have been derived from the effective mean-field evolution of GFT condensate states \cite{Gielen:2016dss,Pithis:2019tvp, Oriti:2016qtz,Gielen:2019kae,Marchetti:2020umh,Jercher:2021bie}, which capture a simple coarse-graining of the underlying quantum gravitational degrees of freedom. The resulting emergent cosmology—formulated relationally with respect to a scalar field clock \cite{Marchetti:2020umh,Marchetti:2024nnk}—exhibits several striking features, including singularity resolution via a quantum bounce \cite{Oriti:2016qtz,Gielen:2019kae,Marchetti:2020umh,Jercher:2021bie,Marchetti:2020qsq,Calcinari:2023sax} and a non-trivial interplay between quantum geometric effects and effective matter dynamics \cite{Ladstatter:2025kgu} (see also Sec.\ \ref{sec:interactinggft} for further discussion). Furthermore, recent extensions beyond the homogeneous regime show that quantum gravitational entanglement can seed cosmological inhomogeneities and lead to modified trans-Planckian dynamics \cite{Marchetti:2021gcv,Jercher:2023kfr,Jercher:2023nxa}.

Importantly, most of the results mentioned above have been obtained in regimes where GFT interactions are negligible. Once interactions are taken into account, they can significantly modify the emergent cosmological dynamics. In particular, for specific classes of interactions, the resulting cosmologies exhibit a late-time de Sitter (dS) phase \cite{deCesare:2016rsf}. These findings—initially derived within highly symmetric single-mode approximations—have been further strengthened by including an additional mode \cite{Oriti:2021rvm,Pang:2025jtk}, revealing that the emergent dark energy component can display phantom-like behavior. More recently, similar indications of dynamical dark energy driven by quantum gravity interactions have also emerged from more realistic GFT condensate models \cite{Ladstatter:2025kgu}.

In this work, we expand on these developments by considering a broad family of interacting generalized mean-field GFT models. We demonstrate that for a large subclass of such models, the cosmological evolution is dominated by an emergent dynamical dark energy component, with the late-time behavior governed by a dS attractor. Moreover, we derive analytic expressions for the time evolution of the effective equation-of-state parameter $w(z)$ across this class of models. We also show that models exhibiting a repulsive dS regime can serve as viable realizations of slow-roll inflation, effectively described by single-field inflation in which the inflaton itself is emergent.

The structure of this Letter is as follows. In Sec.\ \ref{sec:interactinggft}, we introduce interacting mean-field GFT models and outline the general framework. Sec.\ \ref{sec:stability} presents a detailed stability analysis, from both the dynamical systems perspective and the viewpoint of the emergent cosmological evolution. In Sec.\ \ref{sec:ede}, we focus on models featuring a dS attractor and characterize the resulting emergent dynamical dark energy behavior. Sec.\ \ref{sec:slowroll} is devoted to models exhibiting dynamical instabilities, where we demonstrate their viability as slow-roll inflationary scenarios and show how they can be effectively recast as canonical single-field inflation. Finally, Sec.\ \ref{sec:conclusions} contains concluding remarks and outlook.

\section{Interacting Group Field Theories}\label{sec:interactinggft}
The fundamental object in the GFT approach to quantum gravity \cite{Freidel:2005qe,Oriti:2006se,Oriti:2011jm,Gielen:2024sxs,Marchetti:2024tjq} is the GFT field, a (generally) complex-valued field $\varphi:\mathcal{D}\to\mathbb{C}$, where the domain $\mathcal{D}=G\times M$ encodes information about the continuum field content of the theory. More precisely, $\mathcal{D}$ corresponds to the values taken by continuum fields when discretized on a finite element of a $(d-1)$-dimensional spacetime boundary used to define the gravitational path integral. An equivalent characterization of the same domain is that it corresponds to the space of continuum field values {\it at a point} (in the manifold on which they are defined). Thus, the group $G$ carries quantum-geometric data and is typically given by (products of) the Lorentz group $\mathrm{SL}(2,\mathbb{C})$ (or its coverings, or appriate subgroups of the same), while $M$ encodes matter degrees of freedom, so that for $n$ scalar fields one has $M=\mathbb{R}^n$. The specific choice of $\mathcal{D}$ depends on the model under consideration. In the following, we will avoid committing to a particular model. The only element of the usual model building that will play a crucial role is the (quantum) geometric nature of the algebraic data, which will be crucial for the cosmological interpretation of the resulting dynamics.

Given a GFT action $S_{\text{GFT}}[\varphi,\varphi^*]$ (usually constructed such that the GFT partition function generates a discrete matter–gravity path integral), the corresponding mean-field theory is obtained from the averaged quantum equations of motion \cite{Oriti:2016qtz,Marchetti:2020umh,Oriti:2021oux} 
\begin{equation}\label{eqn:meanfield}
    0=\left\langle\delta S_{\text{GFT}}[\hat{\varphi},\hat{\varphi}^\dagger]/\delta\hat{\varphi}^\dagger\right\rangle_\sigma\,,
\end{equation}
and by its hermitian conjugate equation, where $\langle\cdot\rangle_\sigma\equiv \bra{\sigma}\cdot\ket{\sigma}$, and $\ket{\sigma}$ is a state characterized by the macroscopic mean-field $\sigma$, from which we seek to extract the effective continuum gravitational dynamics encoded in the full quantum gravity theory.
Typical choices for $\ket{\sigma}$ are GFT coherent states \cite{Gielen:2013naa,Oriti:2016qtz,Gielen:2019kae,Marchetti:2020umh,Marchetti:2020qsq,Calcinari:2023sax}, which, in the simplest cosmological applications, are restricted so to encode only homogeneous and isotropic data \cite{Oriti:2016qtz}.
In practice, this is implemented through restrictions on the $G$-representation data associated with $\sigma$, together with conditions on its functional dependence on the field content, which is only allowed to involve simple matter data, including those to be used as clock-like variables. In the following, we consider a single minimally coupled, massless, free scalar field $\chi$ serving as a relational clock \cite{Marchetti:2024nnk}, and neglect the possibility of additional matter (since our main purpose is to show how the quantum gravity dynamics itself can produce cosmological acceleration). Decomposing equation \eqref{eqn:meanfield} in $G$-irreps, one obtains \cite{Ladstatter:2025kgu, Oriti:2021oux}
\begin{equation}\label{eqn:generalformeom}
    L_\upsilon[\sigma_\upsilon]+U_\upsilon[\sigma_\upsilon,\sigma^*_\upsilon]=0\,,
\end{equation}
where $\upsilon$ are representation labels, $L_\upsilon$ is a differential operator in relational time of the form\footnote{Depending on the approach taken to implement a relational description \cite{Marchetti:2020umh,Marchetti:2024nnk,Calcinari:2024pek}, the kinetic kernel $L_\upsilon$ may contain a term of the form $-2i\tilde{\pi}_0\partial_\chi$. As the exact form of $L_\upsilon$ is not particularly relevant for the following analysis, we will neglect this term for simplicity.} $L_\upsilon=\partial^2_\chi-E^2_\upsilon$, and $U_\upsilon$ is a potential term. Here, for simplicity, we have made two further assumptions, with respect to the most general case: that the equations for different mean field modes (representation labels) decouple, and that a single interaction term is present. Decomposing $\sigma_\upsilon\equiv \rho_\upsilon e^{i\theta_\upsilon}$, $U_\upsilon[\sigma_\upsilon,\sigma^*_\upsilon]$ is usually assumed to decouple different $\upsilon$-modes and to take the form
\begin{equation}\label{eqn:interactions}
   U_\upsilon[\sigma_\upsilon,\sigma^*_\upsilon] =-\lambda_\upsilon\rho_\upsilon^le^{i[(m+1)\theta_\upsilon+\vartheta_\upsilon]}\,,
\end{equation}
where $l\in \mathbb{N}^+$ is a positive integer, $\lambda_\upsilon\in\mathbb{R}$, and $\vartheta_\upsilon\in\mathbb{R}$. The real and imaginary parts of equation \eqref{eqn:generalformeom} with interactions given by \eqref{eqn:interactions} then take the form (again, derivatives are taken with respect to the clock scalar field variable))
\begin{subequations}\label{eqn:fundamentalequations}
\begin{align}
    0 &= \rho''_\upsilon - \left( \left(\theta'_\upsilon\right)^{2}  + E_\upsilon^{2} \right) \rho_\upsilon - \lambda_\upsilon \cos \left( \vartheta_\upsilon +m \theta_\upsilon \right) \rho_\upsilon^{l}
	\,,
	\\ \label{eq:fullphaseeom}
	0 &= \rho_\upsilon \theta_\upsilon'' + 2 \rho_\upsilon'  \theta_\upsilon'- \lambda_\upsilon \sin\left( \vartheta_\upsilon +m \theta_\upsilon \right)\rho_\upsilon^{l}
    \,.
\end{align}
\end{subequations}
In the existing literature, the parameter $m$ in the above equations has typically been restricted to specific values, corresponding to the so-called pseudotensorial and pseudosimplicial \cite{deCesare:2016rsf,Ladstatter:2025kgu} classes of models.
\paragraph*{Pseudotensorial.}
 For $m=0=\vartheta_\upsilon$, one obtains interactions of the psuedotensorial (PT) type \cite{deCesare:2016rsf}. These arise from an interaction term in the effective mean-field action\footnote{Note that $S_{\text{eff}}$ does not need to coincide with $\braket{S_{\text{GFT}}[\hat{\varphi},\hat{\varphi}^\dagger]}_\sigma$.} $S_{\text{eff}}[\sigma,\sigma^*]$ of the form $\mathcal{V}[\sigma_\upsilon,\sigma^*_\upsilon]=\sum_\upsilon\mathcal{V}_\upsilon[\rho_\upsilon]$, where $\mathcal{V}_\upsilon[\rho_\upsilon]=-\lambda_\upsilon\rho_\upsilon^l$. These interactions have attracted considerable interest due to their high degree of symmetry, and because the corresponding theory space is easy to characterize, which in turn allows for systematic renormalization analyses ~\cite{Carrozza:2013oiy,Carrozza:2016vsq,Pithis:2020kio,Finocchiaro:2020fhl}. Importantly, Landau–Ginzburg techniques show that the mean-field approximation is very robust for these models \cite{Marchetti:2020xvf,Marchetti:2022nrf,Marchetti:2022igl}. From a cosmological perspective (see below and Sec.\ \ref{app:dictionary} for details on the GFT–cosmology correspondence), these models generically exhibit the emergence of a cosmological-constant–like component for $l=5$ once interactions dominate (in a single-mode $\upsilon_0$ scenario) \cite{deCesare:2016rsf}, or a phantom-like dark energy component when two dominant modes ($\upsilon_1$ and $\upsilon_2$) are present \cite{Oriti:2021rvm,Pang:2025jtk}. 
\paragraph*{Pseudosimplicial.}For $m<0$, and more specifically for $m=-l-1$, one obtains interactions of the pseudosimplicial (PS) type \cite{Ladstatter:2025kgu}. These arise from an interaction term in the effective mean-field action of the form $\mathcal{V}[\sigma_\upsilon,\sigma^*_\upsilon]=\sum_\upsilon\mathcal{V}_\upsilon[\sigma_\upsilon]+\text{h.c.}$, with $\mathcal{V}_\upsilon[\sigma_\upsilon]=\gamma_\upsilon\sigma_\upsilon^l$ and $\gamma_\upsilon\in\mathbb{C}$. Their structure is reminiscent of models motivated by simplicial-gravity path integrals. The resulting cosmological dynamics — particularly when interacting matter fields are included — has been studied in detail in \cite{Ladstatter:2025kgu}, revealing the emergence of a dynamical dark-energy component for the first time, together with significant quantum-gravity effects on the continuum scalar-field evolution. These include the appearance of an effective mass term and compatibility constraints on the possible form of the scalar-field potential.
\paragraph*{Generalized mean-field models.} In this work, we extend the above analysis by considering a generalized mean-field model with interactions of the form \eqref{eqn:interactions}, allowing for an arbitrary real value of $m$. Notably, if the linear term $L_\upsilon[\sigma_\upsilon]$ arises from the variation of a real quadratic kinetic term in $\sigma$ and $\sigma^*$, one can immediately observe that a potential of this type cannot originate from a real monomial interaction in $S_{\text{eff}}$ involving only $\sigma$ and $\sigma^*$. This contrasts with the PT and PS cases discussed above. Indeed, for a generic real monomial of the form $\mathcal{V}_\upsilon[\sigma_\upsilon,\sigma^*_\upsilon]=\gamma_\upsilon \sigma_\upsilon^a(\sigma_\upsilon^*)^b+\text{h.c.}$, the variation with respect to $\sigma^*$ produces oscillatory terms with phases $a-b+1$ and $b-a+1$, which cannot reproduce the single phase mode $m+1$ appearing in \eqref{eqn:interactions}, except when $m=0$ or when either $a$ or $b$ vanishes (precisely yielding the PT and PS interactions). By contrast, this matching becomes possible if one considers a non-hermitian (NH) interaction of the form $\mathcal{V}_\upsilon[\sigma_\upsilon,\sigma^*_\upsilon]=\gamma_\upsilon \sigma_\upsilon^a(\sigma_\upsilon^*)^b$ for $a=(l+m+1)/2$ and $b=(l-m+1)/2$. For NH systems, the equations obtained from $\delta S/\delta \sigma_\upsilon = 0$ and $\delta S/\delta \sigma^*_\upsilon = 0$ are not independent (unlike in the hermitian case), and imposing both typically overconstrains the dynamics. One therefore selects a single variational equation as dynamical — in our case, $\delta S/\delta \sigma_\upsilon = 0$ — and discards the other \cite{millington,millington2,Seynaeve:2020oza}.  NH dynamics appear across a wide range of physical settings and are often associated with open-system behavior and dissipation \cite{Ashida:2020dkc}. This is especially compelling in the TGFT context, where the mean-field sector provides a hydrodynamic, coarse-grained description of the underlying quantum-geometric degrees of freedom, for which dissipative effects are generically expected.

Hermitian dynamics can be recovered by considering
\begin{equation}\label{eqn:interactionshermitian}
   \tilde{U}_\upsilon[\sigma_\upsilon,\sigma^*_\upsilon] =-\lambda_\upsilon\rho_\upsilon^le^{i\vartheta_\upsilon}\left[e^{i(m+1)\theta_\upsilon}+c_\upsilon e^{i(1-m)\theta}\right]\,,
\end{equation}
which can be generated by $\mathcal{V}_\upsilon[\sigma_\upsilon,\sigma^*_\upsilon]=\gamma_\upsilon \sigma_\upsilon^a(\sigma_\upsilon^*)^b+\text{c.c}$ for $a=(l+m+1)/2$ and $b=(l-m+1)/2$, $\gamma_\upsilon=-\lambda_\upsilon e^{i\vartheta_\upsilon}/b$ and for an appropriate $c=ae^{2i\vartheta_\upsilon}/b$. The resulting mean-field equations become however more complicated, making the analysis less transparent. For this reason, in the following we will focus only on \eqref{eqn:interactions}, leaving further generalizations to future work.

We will explore the cosmological consequences of the mean-field dynamics \eqref{eqn:generalformeom}. The link between cosmological quantities and GFT data is encoded in the following identity \cite{Oriti:2016qtz}:
\begin{equation}\label{eqn:rhoa}
    V_0a^3=\sum_\upsilon \mathfrak{v}_\upsilon \rho_\upsilon^2\quad\underset{\upsilon_o\text{ dominating}}{\longrightarrow}\mathfrak v_{\upsilon_o}\rho_{\upsilon_o}^2\,,
\end{equation}
where $V_0$ is a fiducial cosmological volume, $\mathfrak{v}_\upsilon$ are eigenvalues of the quantum volume operator on an isotropic volume element (tetrahedron), $\upsilon_o$ is a representation label assumed to dominate the above sum \cite{Oriti:2016qtz} (indications that this regime is dynamically realized can be found in \cite{Gielen:2016uft}). The volume is thus evolving, via the dependence of the density $\rho$, as a function of the clock scalar field $\chi$. From now on, we will focus on this single-representation scenario and drop the subscript $\upsilon_o$ for the sake of notational simplicity. The above expression determines $a(\rho)$, and thus also allows us to completely characterize the physics of the corresponding homogeneous and isotropic universe (see App.\ \ref{app:dictionary} for a summary of the main identities). In view of \eqref{eqn:rhoa}, cosmologically viable solutions are those for which $\rho$ becomes large at late times, corresponding to an expanding universe with large physical volume. From the perspective of the underlying quantum gravity theory, this regime is particularly significant: the density $\rho$ not only dictates the emergent cosmological dynamics but also controls the onset of the classical limit, which is reached for sufficiently large $\rho$ \cite{Marchetti:2020qsq}. In this regime, quantum gravity fluctuations are suppressed\footnote{We emphasize that the suppression of fluctuations in quantum observables does not preclude the appearance of genuinely new quantum gravity effects in the effective classical dynamics, as will be clearly illustrated in the remainder of this work.}, allowing for a direct comparison between GFT mean-field and classical cosmology behavior. 

The rest of this Letter will be devoted to describing the properties of the emergent cosmologies associated with equations \eqref{eqn:fundamentalequations}. While the restriction to a single-representation scenario is a limitation of our analysis, for example compared to \cite{Oriti:2021rvm,Pang:2025jtk}, 
our analysis addresses for the first time more realistic models by studying the cosmological acceleration produced by interactions of the PS and NH type, and in particular the role of the condensate density, thus representing a crucial generalization of previous work.

\section{Asymptotic stability analysis}\label{sec:stability}
When analyzing a dynamical system such as \eqref{eqn:fundamentalequations}, a natural question concerns the existence of attractor solutions at asymptotically late times \cite{kloeden}. In what follows, we therefore concentrate on this asymptotic regime and investigate the stability properties of \eqref{eqn:fundamentalequations}: first from a dynamical systems viewpoint, and subsequently in terms of the associated emergent cosmological evolution. As said, this large density regime corresponds to a large volume universe, resulting from a previous expanding dynamics. 

\subsection{Dynamical stability}
The asymptotic ($\rho\to\infty$) stability properties of \eqref{eqn:fundamentalequations} are more straightforwardly studied by defining
\begin{equation}
    x = \theta\,, \quad y = {\rho'}/{\rho^{3}}\,, \quad z = \theta'\,,
\end{equation}
to recast the dynamics in terms of the following three first-order ODEs:
\begin{subequations}\label{eqn:firstorder}
\begin{align}
    x' &= z
    \,,\\
    y' &= - 3\rho^{2} y^{2} + \frac{z^{2} + E^{2}}{\rho^{2}} + \lambda \rho^{l-3} \cos{\left(\vartheta + m x\right)}
    \,,\label{eqn:yprime}\\
    z' &= - 2 \rho^{2} y z + \lambda \rho^{l-1} \sin{\left(\vartheta + m x\right)}
    \,.
\end{align}
\end{subequations}
This non-autonomous system is characterized by the following family of fixed points:
\begin{subequations}\label{eqn:fixedpoints}
\begin{align}
    \bar{x}_n&= \frac{- \vartheta + n \pi}{m} \,, \quad n \in \mathbb{Z}
    \,,\\
    \bar{y}^{2}_n &= \frac{1}{3} \left( \frac{E^{2}}{\rho^{4}} + \left(-1\right)^{n} \lambda \rho^{l-5}\right)\simeq \frac{(-1)^n\lambda\rho^{l-5}}{3} 
    \,,\\
    \bar{z} &= 0\,.
\end{align}
\end{subequations}
As $\bar{y}_n^2
\ge 0$, we will restrict ourselves to values of $n$ such that $(-1)^n\mathrm{sgn}(\lambda)>0$, so that $\bar{y}^2_n=\bar{y}^2=\vert\lambda\vert\rho^{l-5}/3$. Note that for $l\leq5$ the quantity $\bar{y}$ becomes constant in the limit of large $\rho$, which is required for a consistent fixed point. However, only the marginal value $l=5$ allows for an asymptotically non-vanishing $\rho'$, which is a necessary requirement for a non-trivial cosmological dynamics, see the discussion at the end of the previous section and in App.\ \ref{app:dictionary}. For this reason, from now on, we will fix $l=5$. Analogously, in the following we will assume $\bar{y}$ (and thus $\rho'$ around the fixed point) to be positive, as this corresponds to cosmic expansion.
Linear perturbations $(\xi,\iota, \zeta)$ around $(\bar{x}_n,\bar{y},\bar{z})$ are governed by the following dynamics:
\begin{equation}\label{eqn:linearized}
    \begin{pmatrix}
         \xi' \\ \iota' \\ \zeta'
    \end{pmatrix}
    =\begin{pmatrix}
        0 & 0 & 1 \\ 
        0 & - 6 \rho^{2} \bar{y} & 0 \\
        3\rho^{4}\bar{y}^2m& 0 & - 2 \rho^{2} \bar{y}
    \end{pmatrix}
    \begin{pmatrix}
        \xi \\ \iota \\ \zeta
    \end{pmatrix}\,.
\end{equation}
The time dependence of the above matrix forbids a trivial eigenvalue-stability analysis, and suggests instead to derive instead explicit solutions to equation \eqref{eqn:linearized}.
We immediately find
\begin{equation}\label{eqn:firstordery}
    \iota=(\rho/\rho_0)^{-6}\,,
\end{equation}
with $\rho_0$ an appropriate integration constant. The $y$-direction is therefore stable, in the large $\rho$ limit. For $\xi$, one can note that its equation is in fact an Euler-type ODE, of the form
\begin{equation}
    \tau^2\frac{\diff^2}{\diff\tau^2}\xi(\tau)-\tau\frac{\diff}{\diff\tau}\xi(\tau)-\frac{3m}{4}\xi(\tau)=0\,,
\end{equation}
where $\tau=\rho^{-2}$. The Euler ansatz $\xi(\tau)=\tau^k$ leads to $k=1\pm \mu$, $\mu=\sqrt{1+3m/4}$. Depending on the value of $1+3m/4$ one thus finds different solutions:
\begin{equation}\label{eqn:deltax}
   \rho^2 \xi(\rho)=\begin{cases}
        A\rho^{-2\mu}+B\rho^{2\mu}\,,\quad &1+3m/4>0\\
        A+B\log\rho\,,\quad &1+3m/4=0\,,\\
        A\cos\Phi(\rho)+B\sin\Phi(\rho)\,,\quad &1+3m/4<0\,,
    \end{cases}
\end{equation}
where $\Phi(\rho)\equiv \beta\log\rho$ and $\beta\equiv 2\sqrt{-1-3m/4}$ for $1+3m/4<0$. The associated $\zeta$ is then given by
\begin{equation}\label{eqn:deltaz}
    \zeta(\rho)=-2\bar{y}\begin{cases}
        A(1+\mu)\rho^{-2\mu}+B(1-\mu)\rho^{2\mu}\,,
        \\
        (A+B/2)+B\log\rho\,, 
        \\
        c_A\cos\Phi(\rho)+c_B\sin\Phi(\rho)\,, 
        \\
    \end{cases}
\end{equation}
for the three cases above, respectively, and where $c_A=A-B\beta/2$, $c_B\equiv B+A\beta/2$. From the above expressions, it is clear that for $1+3m/4\ge 0$ the fixed point is asymptotically unstable, as perturbative solutions grow larger with $\rho$. For $1+3m/4<0$, instead, the fixed points in \eqref{eqn:fixedpoints} are not attractors in a strong sense. Rather, they are (local) \emph{weak forward attractors} \cite{kloeden,chepyzhov2002attractors,Vishik_2011}, meaning that for every bounded set $B\subset \mathcal{X}$, with $\mathcal{X}$ the space of initial data for $(\xi,\zeta)$  
\begin{equation}
    \mathrm{dist}_{\mathrm{w}}(U(\rho,\rho_0)B),\{0\})\to 0\,,\\ \quad \rho\to \infty\,,
\end{equation}
where $\mathrm{dist}_{\mathrm{w}}(A,B)$ is the weak distance between $A$ and $B$, $U(\rho,\rho_0)= U(\chi,\chi_0)$ is the evolution operator satisfying $U'=SU$, with $U(\chi_0,\chi_0)=\mathbb{I}$, and where $\tilde{S}$ is the non-diagonal stability matrix obtained from $S$ by removing the second row and column (note that the $y$-direction can be disregarded in this analysis as it is trivially stable). A proof of this statement can be found in App.\ \ref{app:[proof]}. In other words, the above equation implies that any perturbation $(\xi,\zeta)$ is weakly indistinguishable from $\{0\}$ as $\rho\to\infty$, meaning that the system will effectlively stabilize at (one of the) fixed points \eqref{eqn:fixedpoints}. 

\subsection{Cosmological stability}\label{sec:cosmologicalstability}
Above, we have seen that the dynamical system characterized by \eqref{eqn:firstorder} has different stability properties depending on the value of $1+3m/4$. However, it is important to emphasize that whether the fixed points \eqref{eqn:fixedpoints} are attractors from the perspective of the dynamical system \eqref{eqn:firstorder}, and whether the resulting cosmological dynamics exhibits attractor behavior are two different questions. To see this, let us recall that it is only the $y$ behavior which determines the emergent cosmological dynamics.
In turn, this offers a very clear cosmological interpretation for the fixed points \eqref{eqn:fixedpoints}. Indeed, from equations \eqref{eqn:eosparameter} and \eqref{eqn:friedmann}, we have
\begin{subequations}\label{eqn:cosmologicalparameters}
\begin{align}\label{eqn:yprimew}
    w&=-[1+y'/(\rho^2 y^2)]\,,\\
    H^2&=\frac{4\pi_\chi^2}{9\mathfrak{v}^2}y^2\,,
\end{align}
\end{subequations}
where $w$ is \textit{defined} as the equation of state parameter of the fluid that effectively sources the emergent cosmological dynamics, $\pi_\chi$ is the $\chi$-momentum in cosmic time gauge and $\mathfrak{v}\equiv\mathfrak{v}_{\upsilon_0}$ is the volume eigenvalue associated to the label $\upsilon_0$ (see App.\ \ref{app:dictionary} for more details). At the fixed point $\bar{y}$, we have from the above equations\footnote{From the quantum gravity theory the constant determining the dS behavior with respect to the relational clock depends only on $\lambda$ and $\mathfrak{v}$, and is obtained from the above expression by setting $\pi_\chi = 1$. In order to make contact with standard cosmology formulations, we work in cosmic gauge and reintroduce $\pi_\chi$.}
\begin{subequations}
\begin{align}
    \bar{w}&=-1\,,\\
    \bar{\Lambda}&=\vert\lambda\vert\frac{4\pi_\chi^2}{9\mathfrak{v}^2}\,,
\end{align}
\end{subequations} 
so that the universe is dominated by an emergent cosmological constant proportional to the strength of the underlying quantum gravity interactions $\lambda$. Note that since both $\bar{w}$ and $\bar{\Lambda}$ only depend on $\bar{y}$, they are independent of $n$ (and $m$), and thus on the exact value of the fixed point. Finally, let us remark that such an accelerating phase can only be obtained for the case $l=5$ (as already pointed out below equations \eqref{eqn:fixedpoints}), confirming indications already obtained in the PT case \cite{deCesare:2016rsf,Oriti:2021rvm,Pang:2025jtk}.

Importantly, irrespective of the initial configuration, any cosmological state will eventually be driven towards the above de Sitter solution. In fact, equation \eqref{eq:fullphaseeom} shows that the $\theta$–dynamics originates from both a forcing term, $-\lambda\sin(\vartheta+m\theta)\rho^{l-1}$ and a dissipative term, $2\theta'\rho'$. Provided that $\rho' > 0$—a necessary condition for a cosmological interpretation of the solution—$\theta$ dissipates energy, and as $\rho$ grows it can no longer overcome the local potential barrier. Consequently, it is driven to the nearest fixed point. Since all fixed points lead to identical physical properties, the cosmology in the vicinity of such a fixed point is effectively dominated by a dynamical dark energy component with a non-constant equation-of-state parameter $w$. Whether this fixed point is attractive (stable) or repulsive (unstable) depends on the sign of the $\theta$–forcing term. Near the fixed points, the term $-\lambda\sin(\vartheta+m\theta)\rho^{l-1}$ is repulsive for $m>0$ and attractive for $m\le 0$. Therefore, it is the value of $m$ that determines the stability of the emergent dS phase, rather than the quantity $1+3m/4$.

This can be confirmed by explicitly computing the dynamics of perturbations in $y$ around $\bar{y}$. As linear perturbations in $y$ are heavily suppressed at late times (cfr.\ equation \eqref{eqn:firstordery}), it is natural to look for corrections in $y$ at second order, triggered by perturbations of the form \eqref{eqn:deltax} and \eqref{eqn:deltaz}. At that order, we have (see App.\ \ref{sec:computations} for explicit computations)
\begin{equation}
    {\vert y-\bar{y}\vert}/{\bar{y}}\sim\mathcal{O}(\xi^2)\,,
\end{equation}
for any value of $m$. Analogously, at the same perturbative order, $y'/(\rho^2\bar{y}^2)\sim\mathcal{O}(\vert y-\bar{y}\vert/\bar{y})$. From equation \eqref{eqn:deltax}, we see that 
only for $\mu-1>0$, and thus $m>0$, $\xi$ has a growing mode\footnote{Note that the dominant modes decrease at most as $\rho^{-4}$, and hence are asymptotically dominating over the first order perturbations \eqref{eqn:firstordery}.}. Thus, for $m\le 0$, $y\to\bar
{y}$ and $y'/(\rho^2y^2)\to 0$, meaning that the emergent dS phase is indeed a cosmological attractor for these models.

The validity of these arguments can be confirmed by a numerical analysis.
\begin{figure}[htbp]
\centering
\includegraphics[width=.45\textwidth]{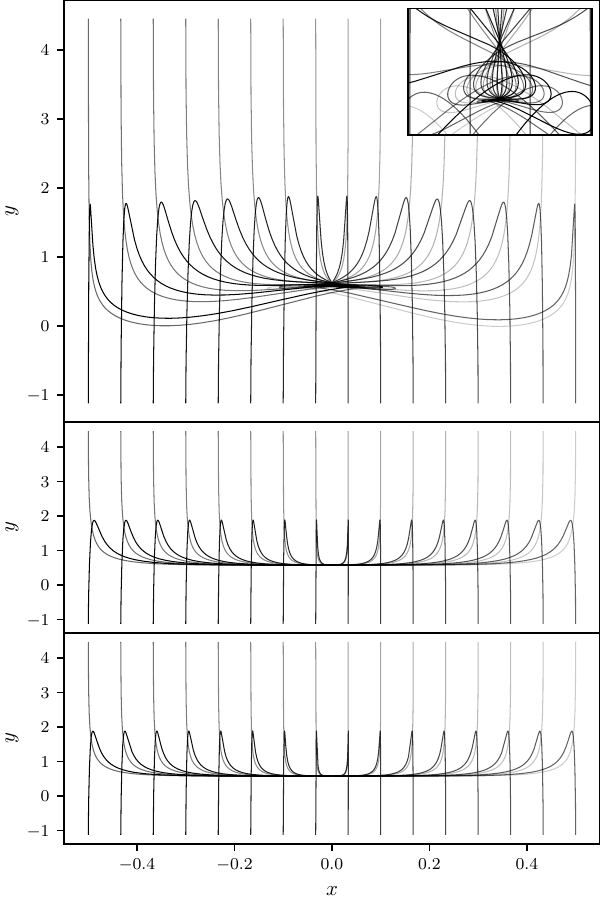}
\caption{\qmarks{Phase space} diagrams showing the presence of a dS attractor for $m=-6$ (top), $m=-4/3$ (middle), and $m=-1$ (bottom), together with a close-up of the self-similar center for $m=-6$. ``Wobbly'' lines are likely due to lack of numerical precision at later times. The microscopic parameters and initial conditions are the same for each plot, and are chosen to be $E=1000.0$, $\vartheta=0.0$, $\lambda=1.0$, $l=5.0$, and $\rho_{0}=15.0$, $\theta'_0=1.0$, respectively.\label{fig:attractor_ps}}
\end{figure}
In Fig.\ \ref{fig:attractor_ps}, numerical solutions of \eqref{eqn:fundamentalequations} show how arbitrary initial states spiral towards the dS attractor at $\bar{y}>0$ and $\bar{x} =0$ for any $m<0$, regardless of the sign of $1+3m/4$, as expected from the above analysis. As it can be seen in the top subplot ($m=-6$), the center is self-similar as the plot repeats itself around the fixed point. It should be noted that the two lower subplots ($m=-4/3$ and $m=-1$) actually cover less of the \qmarks{phase space} than the top one, since the periodic boundaries in $x$ are given by $[-\pi/m,\pi/m]$, which here is the smallest for $m=-6$. We synchronized the subplots and initial conditions for illustrative purposes.
\begin{figure}[htbp]
\centering
\includegraphics[width=.45\textwidth]{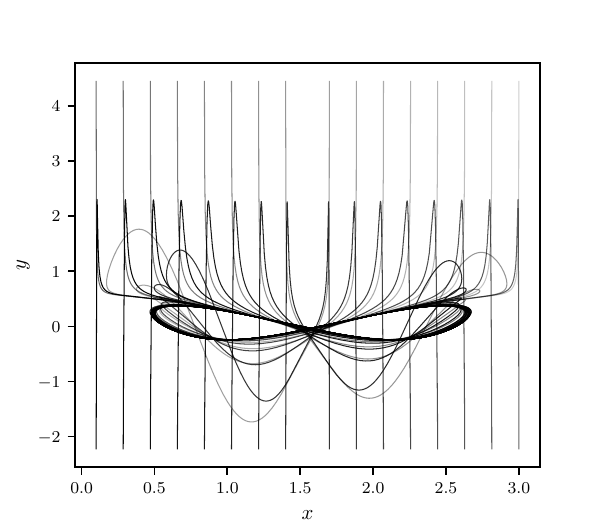}
\caption{\qmarks{Phase space} diagram showing the repellent nature of the dS fixed point for $m=2>0$. Independently of the initial conditions considered, the system establishes orbits bounded by the dS fixed points $\bar{x}_0=0$ and $\bar{x}_2=\pi$ and centered around $\bar{x}_1=\pi/2$. Microscopic parameters and initial conditions are as in Fig.\ \ref{fig:attractor_ps}.
\label{fig:attractor_unstable}}
\end{figure}

On the other hand, Fig.\ \ref{fig:attractor_unstable} illustrates the dynamics for $m=2>0$. As discussed above, the system is initially driven toward one of the (real, $n$ even) fixed points, $\bar{x}_0=0$, $\bar{x}_2=\pi$, for any initial $x_0\in [\bar{x}_0,\bar{x}_2]$. However, these fixed points are repulsive, and once the system enters their vicinity, it is pushed away, resulting in sustained oscillations centered around $\bar{x}_1=\pi/2$ and bounded by $\bar{x}_0$ and $\bar{x}_2$. The same behavior is clearly established for any $x_0\in[\bar{x}_n,\bar{x}_{n+2}]$ for arbitrary even $n$, and thus for arbitrary initial conditions.

In the following, we will study in detail the physical properties of the emergent cosmologies in two cases $m\le0$ (Sec.\ \ref{sec:ede}) and $m>0$ (Sec.\ \ref{sec:slowroll}) around the dS fixed point(s).

\section{Emergent dynamical dark energy}\label{sec:ede}
As shown above, for $m \le 0$ the emergent dS phase acts as a cosmological attractor. In this section, we precisely characterize the late–time behavior of the cosmological parameters $\Lambda(z)$ and $w(z)$. We first derive these results perturbatively and analytically (Sec.\ \ref{sec:edepert}), and subsequently validate them against non-perturbative numerical computations (Sec.\ \ref{ssec:dSnumerics}). 
\subsection{$\Lambda$- and $w$-dynamics}\label{sec:edepert}
As discussed in Sec.\ \ref{sec:cosmologicalstability}, equations \eqref{eqn:cosmologicalparameters} relate the evolution of the emergent dark energy parameters to the dynamics of $y$. The latter can be explicitly computed at second order in $\xi$ (see App.\ \ref{sec:computations}). More precisely, given such a perturbative solution $y(\rho)$, and defining 
\begin{subequations}
\begin{align}
    w(z)&\equiv-1+\delta w(z)\\
    \Lambda(z)&\equiv\bar{\Lambda}(1+\delta\ell(z))\,,
\end{align}
\end{subequations}
one can obtain $\delta w(z)$ and $\delta\ell(z)$ by using equations \eqref{eqn:cosmologicalparameters} and $\rho^2=[(1+z_q)/(1+z)]^3$, with $z_q\equiv z(\rho=1)$. Below we provide analytic expressions for $\delta\ell(z)$ and $\delta w(z)$ depending on the value of $1+3m/4$ with $m\le 0$.
\paragraph*{$1+3m/4<0$.} For $1+3m/4<0$ (which includes the PS case $m=-6$), we can use equations \eqref{eqn:yofg}, \eqref{eqn:g} and \eqref{eqn:yprimecase3} to obtain
\begin{subequations}
\begin{align}
    \delta \ell(z)&=\left(\frac{1+z_q}{1+z}\right)^{-6}[\delta \ell_0+\delta\ell_1\cos2\Phi(z)\nonumber\\
    &\qquad\qquad\qquad\quad+\delta\ell_2\sin 2\Phi(z)]\\ 
\label{eqn:deltaw1}
    \delta w(z)&=\left(\frac{1+z_q}{1+z}\right)^{-6}[\delta w_0+\delta w_1\cos2\Phi(z)\nonumber\\&\qquad\qquad\qquad\quad+\delta w_2\sin2\Phi(z)]\,,
\end{align}
\end{subequations}
where $\Phi(z)=(3\beta/2)\log[(1+z_q)/(1+z)]$, and where $\delta\ell_i=\delta\ell_i(A,B;m)$, $\delta w_i=\delta w_i(A,B;m)$ for $i=0,1,2$ depend on $m$ and are quadratic functions of the initial $\xi$-data $(A,B)$. The explicit form of $\delta \ell_i,\delta w_i$ can be obtained by direct comparison with equations \eqref{eqn:yofg}, \eqref{eqn:g}, \eqref{eqn:yprimecase3} and \eqref{eqn:deltawk}, see e.g.\ Fig.\ \ref{fig:deltaw} (top) for $m=-6$. Note that the maximum value attained by the factor in square brackets in equation \eqref{eqn:deltaw1} is given by $\delta \tilde{w}_{\max}\equiv \delta w_0+\sqrt{\delta w_1^2+\delta w_2^2}$. Rewriting $\delta \tilde{w}_{\max}$ as a function of $A$, $B$ and $m$, one can show that, for any real value of $A$ and $B$, 
\begin{equation}
    \delta w(z)\le 0\,,\qquad\longleftrightarrow \qquad m\le -6\,.
\end{equation}
Thus, the emergent dark energy exhibits generic phantom behavior for 
$m\le -6$, with the PS case $m=-6$ marking the boundary between the phantom and non-phantom regimes.

\paragraph*{$1+3m/4\ge 0$.} In this regime, corresponding to $0\le \mu\le 1$ for $m\le 0$ (which thus includes the PT case), we can use equations \eqref{eqn:yofxi} and \eqref{eqn:yprimeothercases} to obtain
\begin{equation}
    \delta w(z)=2(1-\mu)\delta\ell(z)\,.
\end{equation}
Moreover, assuming the growing mode in equations \eqref{eqn:yprimeothercases} and \eqref{eqn:deltax} to be dominating, we obtain
\begin{subequations}\label{eqn:deltaw2}
\begin{align}
    \delta w(z)&=\delta w_0\left(\frac{1+z_q}{1+z}\right)^{6(\mu-1)} \quad 0<\mu\le 1\,,\label{eqn:deltawcase1}\\
    &=9\delta w_0\left(\frac{1+z_q}{1+z}\right)^{-6}\log^2\frac{1+z_q}{1+z}\,,\quad\mu=0\,,\label{eqn:deltawcase2}
\end{align}
\end{subequations}
where $\delta w_0=\delta w_0(B;m)=2m^2B^2(1-\mu)y_1(m)$ depends on the initial condition on the growing mode, and where $y_1(m)$ is defined by \eqref{eqn:y1}, see e.g. Fig.\ \ref{fig:deltaw} for $m=-4/3$ (middle) and $m=-1$ (bottom). Note that within the relevant regime $m<0$, $\delta w_0<0$ within the region $2(3-2\sqrt{6})/3\le m<0$, implying in particular that for $m=-4/3$ (and thus in the case \eqref{eqn:deltawcase1}) the dynamics cannot be phantom-like.

To summarize, for all values of $m\ge 0$, the time-dependent component of the emergent dark-energy parameters is characterized by a suppression term of the form $[(1+z_q)/(1+z)]^{-6}$, which is respectively modulated by trigonometric-logarithmic, pure logarithmic and mildly growing monomial functions of $[(1+z_q)/(1+z)]$ depending on the value of $1+3m/4$ (see Fig.\ \ref{fig:deltaw}). The qualitative $z$-dependence of these parameters therefore differs from the one obtained in \cite{Oriti:2021rvm,Pang:2025jtk} (and proportional to $(1+z)^3$) for the two-mode PT case. Moreover, the equation of state parameter shows generic (initial-conditions independent) phantom-like behavior at late times for $m\le -6$ (thus including the PS case) and $2(3-2\sqrt{6})/3\le m<0$, similarly to the findings of \cite{Oriti:2021rvm,Pang:2025jtk} in the two-mode PT case.

\paragraph*{Impact on the Hubble parameter.}
The dependence of the effective equation of state parameter on relational time affects the cosmic expansion history, and hence the estimation of the Hubbble parameter today, $H_0$. Indeed, one can write
\begin{equation}\label{eqn:deltaH}
    \frac{\delta H}{H}\simeq 3\int_0^z\frac{\delta w(z')}{1+z'}\diff z'\,,
\end{equation}
where $\delta w(z)$ for $m\le 0$ is given by equations \eqref{eqn:deltaw1} and \eqref{eqn:deltaw2} for $1+3m/4<0$ and $1+3m/4\ge0 $, respectively.
As we see from equation \eqref{eqn:deltaH}, the sign of $\delta H/H$ depends crucially on the sign of $\delta w(z)$. In turn, this determines the change in $H_0$ as follows \cite{Heisenberg:2022gqk}. 

Given a change in the expansion history that produces a corresponding change in the estimation of $H_0$, any cosmological observable $\mathcal{O}$ changes as
\begin{equation}
    \frac{\Delta\mathcal{O}(z)}{\mathcal{O}(z)}=I_{\mathcal{O}}(z)\frac{\delta H_0}{H_0}+{\int_0^\infty} {\frac{\diff z'}{1+z'}}R_{\mathcal{O}}(z,z')\frac{\delta H(z')}{H(z')}\,.
\end{equation}
One can then determine the induced change on $H_0$ by determining an observable whose value at a certain redshift $z_\star$, $\mathcal{O}(z_\star)\equiv \mathcal{O}_\star$ should not be affected by the modified cosmological model, $\Delta \mathcal{O}_\star=0$, leading to
\begin{equation}
    \frac{\delta H_0}{H_0}=-{\int_0^\infty} {\frac{\diff z'}{1+z'}}\frac{R_{\mathcal{O}}(z_\star,z')}{I_{\mathcal{O}}(z_\star)}\frac{\delta H(z')}{H(z')}\,.
\end{equation}
The selection of suitable observables is not entirely straightforward within the simplified framework adopted here. Standard CMB priors—such as the acoustic scale $\theta_\star$ or the shift parameter $R_\star$—cannot be employed due to the absence of radiation in this model. Nevertheless, the above expressions, together with equation \eqref{eqn:deltaH}, provide a direct means to compute the change in $H_0$ once these results are embedded in a realistic cosmological setting or, preferably, when radiation degrees of freedom are coupled to the underlying GFT dynamics.

\subsection{Numerics}\label{ssec:dSnumerics}
As in the previous section, the above results can be corroborated by numerical analysis.

\begin{figure}[htbp]
\centering
\includegraphics[width=.45\textwidth]{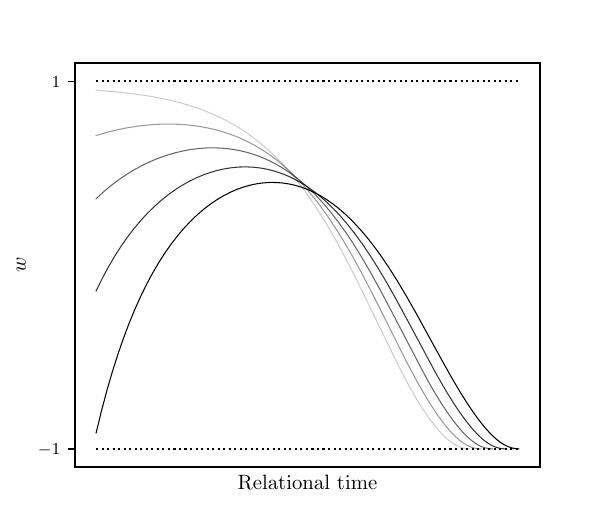}
\caption{Different initial $w$ converging to $w=-1$. Microscopic parameter and initial conditions are $E=1000.0$, $\vartheta=0.0$, $\lambda=1.0$, $m=-6$, $l=5.0$, and $\rho_{0}=15.0$, $\theta_0=1.0$, $\theta'_0=1.0$, respectively \label{fig:attractor_1}}
\end{figure}

First, Fig.\ \ref{fig:attractor_1} shows the evolution of the equation of state parameter $w$ for $m=-6$ and for a set of different initial values for $w$ within the cosmologically interesting range $[-1,1]$, illustrating once again the presence of a late time dS attractor. 

\begin{figure}[htbp]
\centering
\includegraphics[width=.45\textwidth]{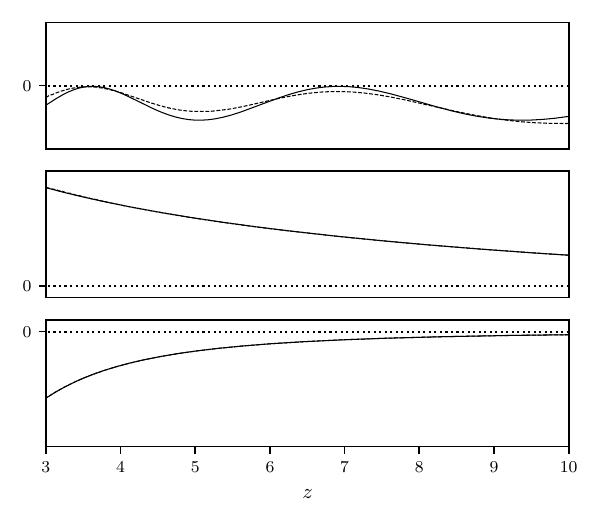}
\caption{Numerically obtained $\delta w(z)[(1+z_q)/(1+z)]^6$ (dashed lines), for $m=-6$ (top), $m=-4/3$ (middle), and $m=-1$ (bottom) respectively, fitted with the analytic solutions of Sec.\ \ref{sec:edepert} (solid lines). Microscopic parameters and initial conditions are $E=1.0$, $\vartheta=0.0$, $\lambda=1.0$, $l=5.0$, and $\rho_{0}=15.0$, $\rho'_{0} = E \rho_{0}$ $\theta_0=1.0$, $\theta'_0=1.0$, respectively}
\label{fig:deltaw}
\end{figure}

In Fig.\ \ref{fig:deltaw}, the numerical results for $\delta w (z)[\left(1 + z_{q}\right)/\left(1 + z\right)]^6$ are compared with fits by the expected analytic behavior of equations \eqref{eqn:deltaw1} and \eqref{eqn:deltaw2}. Note that, when fitting the numerical data, $z_q$ should not be treated as a free parameter. Indeed, equations \eqref{eqn:deltaw1} and \eqref{eqn:deltaw2}, as well as the numerical form of $w$, are functions of $\rho$ and can therefore be fitted without any reference to $z_q$. The parameter $z_q$ is only required to express the evolution in terms of $z$ via equation \eqref{eqn:rhotoz}. This situation would be different if we were fitting real observational data $w_{\text{obs}}(z)$, which are naturally expressed in terms of redshift: in that case, \eqref{eqn:deltaw1} and \eqref{eqn:deltaw2} would require $z_q$ as an additional parameter to fit, encoding the density of quantum gravity atoms today. To mimic this realistic setup, we still perform the fit in terms of $z$, but we (arbitrarily) fix $z_q=\rho_{\mathrm{end}}^{2/3}-1$ (see equation \eqref{eqn:nandz}) by identifying today with the endpoint of the numerical integration, characterized by $\rho=\rho_{\mathrm{end}}$.
Furthermore, the plots and fits have been restricted to a limited range of $z\in[3,10]$ for the following reasons. At large $z$, the perturbative regime breaks down, so one should not expect \eqref{eqn:deltaw1} and \eqref{eqn:deltaw2} to remain valid. At small $z$, numerical precision decreases, introducing substantial noise. The ranges on all $y$-axes in Fig.\ \ref{fig:deltaw} have been omitted, since the absolute scales depend arbitrarily on the choice of initial conditions.

More precisely, the $m=-6$ model is fitted using \eqref{eqn:deltaw1}, with $A$, $B$, and $\beta$ as fit parameters. The values of $A$ and $B$ depend on the details of the numerical solution and are therefore not physically meaningful. However, the best-fit value of $\beta$ is $\beta_{\text{bf}} = 3.9 \simeq \sqrt{14}$, which is approximately consistent with the expected analytic prediction.
The $m=-4/3$ model is fitted using $\left(\delta w_c + \delta w_0 \log{\left(\left(1 + z_{q}\right)/\left(1 + z\right)\right)}\right)^{2}$, obtained by combining equations \eqref{eqn:yprimeothercases} and \eqref{eqn:deltax}. This is necessary because \eqref{eqn:deltawcase2} includes only the dominant contribution to $\delta w(z)$ under the assumption of extremely large $z_q$, an approximation unsuitable for comparison with numerical solutions, since such large values of $z_q$ (and hence $\rho_{\text{end}}$) were not reached numerically. As before, the best-fit values of $\delta w_c$ and $\delta w_0$ do not convey relevant information, but the quality of the fit is clearly excellent.
Lastly, the $m=-1$ model is fitted using \eqref{eqn:deltawcase1}, with $\delta w_0$ and $6\mu$ as fit parameters. As in the previous cases, $\delta w_0$ conveys no relevant information, while the best-fit value of the exponent is $6\mu_{\text{bf}} = 3.03$, consistently with the analytic prediction $6\mu=3$.

In summary, Fig.\ \ref{fig:deltaw} shows very good agreement between the full, non-perturbative numerical solution and the perturbative analytic ones in regimes in which both types of solutions are reliable. In particular, it confirms the emergence of a late-times dynamical dark energy with phantom behavior for $m=-6$ and $m=-1$, which lie in the analytically determined regions $m\le -6$ and $2(3-2\sqrt{6})/3\le m<0$.

\section{Emergent Slow-roll inflation}\label{sec:slowroll}
In Sec.\ \ref{sec:stability}, we showed that for $m<0$ the emergent dS phase is a cosmological attractor. Conversely, for $m>0$ ($\mu>1$) we found that $y$–instabilities arise at second order due to perturbative growth in $\xi$. From the viewpoint of the emergent cosmology, when the universe is in an accelerating state—namely in the vicinity of the fixed points \eqref{eqn:fixedpoints}—these dynamical instabilities will ultimately trigger an exit from acceleration. Depending on the duration of this accelerating period before instability takes over, such solutions may represent realistic slow-roll inflationary scenarios. For this to give a viable cosmological model, the acceleration must occur in the early universe and at comparatively smaller density/volume. Since the onset of acceleration is controlled by the interaction strength $\lambda$, viable inflationary models correspond to comparatively larger coupling constants than those associated with models that instead yield dynamical dark energy at late times.

To study this inflationary phase in detail, we first characterize it analytically within a perturbative framework (Sec.\ \ref{subsec:SRperturbative}), and then support these results with numerical analysis (Sec.\ \ref{sec:inflationnumerics}). We conclude with preliminary (analytic and numerical) considerations on the ensuing post-inflationary dynamics (Sec.\ \ref{sec:postinflation}).
\subsection{Perturbative and slow-roll analysis}\label{subsec:SRperturbative}
To characterize the instabilities around the dS solution and to make contact with inflationary physics, it is convenient to introduce the slow-roll parameters \cite{Martin:2013tda} $\epsilon_{n+1}\equiv \diff \log\vert\epsilon_n\vert/\diff N$ for $n\ge 0$, where $\epsilon_0=H_{\text{in}}/H$ and $N\equiv \log (a/a_{\text{in}})$ is the number of e-folds. Recall that by definition $\epsilon_1(z)=3\delta w(z)/2$. In turn, for $m<0$ one can still use equations \eqref{eqn:yprimeothercases} to obtain a $\delta w(z)$ of the same functional form as \eqref{eqn:deltawcase1}. Combining equations \eqref{eqn:nandz} to obtain $z(N)$, we can then write
\begin{subequations}
\begin{align}
\epsilon_1(N)&=\varepsilon_1\left(\frac{1+z_q}{1+z_{\text{end}}}\right)^{6(\mu-1)}e^{-6(\mu-1)(N_{\text{end}}-N)}\,,\\
    \epsilon_2(N)&=6(\mu-1)\,,\\
    \epsilon_n(N)&=0\,,\qquad n\ge 3\,,
\end{align}
\end{subequations}
which truncates the slow-roll series. In the above equations, $\varepsilon_1=\varepsilon_1(B;m)=3\delta w_0(B;m)/2>0$ for $m>0$, and $z_{\text{end}}=z(N_{\text{end}})$ is the redshift at which inflation ends.
By construction, at $N_{\text{end}}$ we must have $\epsilon_1(N_{\text{end}})=1$. Since $\varepsilon_1(B;m)$ is not a large quantity, contrarily to $\rho_{\text{end}}^{2/3}=[(1+z_q)/(1+z_{\text{end}})]$, this can only be achieved if $\epsilon_2>0$, which is indeed guaranteed in the $m>0$ case we are considering\footnote{Note that the initial condition $\vert\epsilon_1(0)\vert\ll 1$ is automatically achieved in this case.}. Nonetheless, even in this regime, the graceful exit condition implies $\varepsilon_1(B;m)=\rho^{-2\epsilon_2/3}_{\text{end}}$, which would require significant fine tuning in $B$, unless $\epsilon_2\ll 1$. In this regime, characterized by $\mu-1\ll1$, (or, equivalently, $ m\ll 1$, and thus $\varepsilon_1(B;m)\simeq 5B^2m^2/32$), the universe undergoes a canonically slow-roll inflationary phase.
 
As argued in \cite{Mukhanov:2013tua}, given $\epsilon_1(N)$ one can always introduce a fictitious self-interacting scalar field $\phi$ driving the inflationary dynamics. 
Using equations \eqref{eqn:emergentscalar}, and defining $\tilde{\phi}=\pm \epsilon_2\phi/(2\sqrt{2}M_{\text{pl}})$, this emergent inflaton and its associated potential are given by
\begin{subequations}
\begin{align}\label{eqn:inflaton}
    \tilde{\phi}(N)-\tilde{\phi}_0&=e^{-\epsilon_2(N_{\text{end}}-N)/2}\,,\\
     \frac{V(\tilde{\phi})}{V_0}&=\left[1-\frac{(\tilde{\phi}-\tilde{\phi}_0)^2}{3}\right]e^{-2(\tilde{\phi}-\tilde{\phi}_0)^2/\epsilon_2}\,,\label{eqn:inflatonpot}
\end{align}
\end{subequations}
so that $\tilde{\phi}(0)\simeq\tilde{\phi}_0$, and $\tilde{\phi}(N_{\text{end}})\simeq \tilde{\phi}_0+1$. The potential introduced above belongs to the Gaussian class; notably, \cite{Ladstatter:2025kgu} showed that such Gaussian effective scalar potentials meet strict quantum-gravity constraints ensuring a consistent emergence of classical dynamics in both the matter and geometric sectors. Moreover, its form is reminiscent of typical SUGRA models where $V(\tilde{\phi})\sim e^{K}f[W]$, with $K$ the Kahler potential and $f$ a functional of the superpotential $W$ \cite{Ketov:2014qha, Ellis:2014rxa}. Finally, note that one can use equations \eqref{eqn:inflaton} and \eqref{eqn:deltax} together with the exit condition $\epsilon_1(N_{\text{end}})=1$ to suggestively write
\begin{subequations}
\begin{align}
    \xi(\tilde{\phi)}&=B\rho_{\text{end}}^{\epsilon_2/3}(\tilde{\phi}-\tilde{\phi}_0)\,,\\
    \epsilon_1(\tilde{\phi})&=(\tilde{\phi}-\tilde{\phi}_0)^2\,,
\end{align}
\end{subequations}
identifying (modulo unimportant prefactors) the GFT phase degree of freedom $\xi=x-\bar{x}_n$ as the inflaton driving the emergent inflationary dynamics.
\subsection{Numerics}\label{sec:inflationnumerics}
\begin{figure}[htbp]
\centering
\includegraphics[width=.45\textwidth]{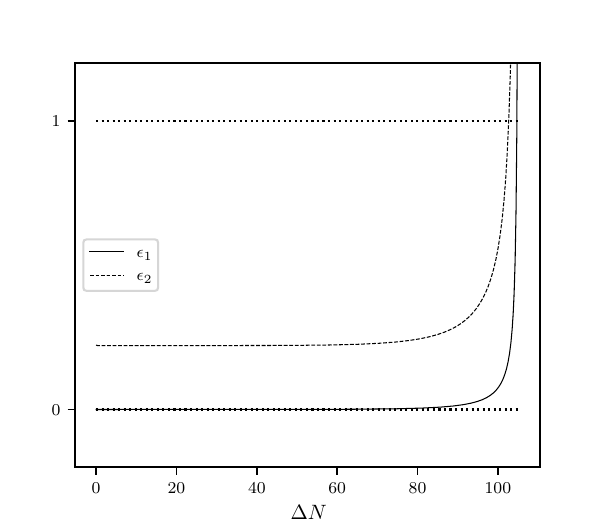}
\caption{Slow roll parameters for $m=0.1$ starting near $0$ and eventually exiting the regime. Dotted lines mark $0$ and $1$. Microscopic parameters are $E=10.0$, $\vartheta=0.0$, $\lambda=10.0$, $m=0.1$, $l=5.0$, while initial conditions are determined by equations \eqref{eqn:unstableansatz} and \eqref{eqn:y1} with $\theta_0=x_0=0.1$. \label{fig:slowroll_1}}
\end{figure}
The above perturbative analysis can be confirmed by non-perturbative numerical solutions of equations \eqref{eqn:xdynamics}. Fig.\ \ref{fig:slowroll_1} shows the evolution of the first two slow-roll parameters starting from $x(N=0)=0.1$ for $m=0.1$, $\lambda = 10$, and with perturbative initial conditions. 
Testing shows that smaller values of $m$ greatly increase $N_{\text{end}}$, as expected. Similarly, smaller initial phase $x(N=0)$, corresponding to an initial state closer to dS, also delays the escape from the unstable fixed point and leads to a larger $N_{\text{end}}$. However, a significant amount of fine-tuning is required to achieve large amounts of expansion in this case. 
\begin{figure}[htbp]
\centering
\includegraphics[width=.45\textwidth]{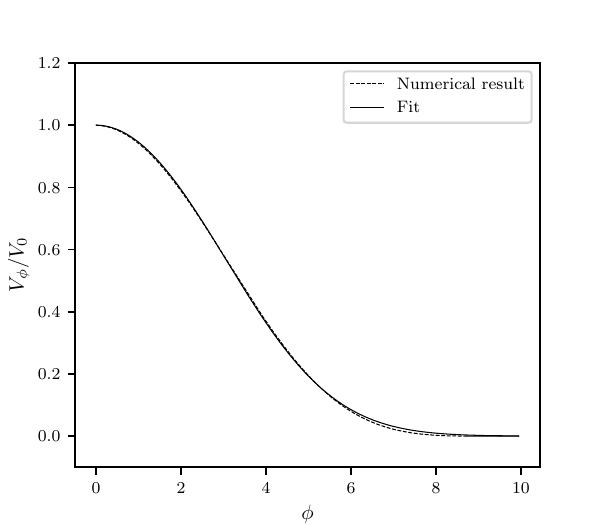}
\caption{Effective potential, obtained  with the same parameters and initial conditions of Fig.\ \ref{fig:slowroll_1}, fitted with $(1-a \phi^{2}) \ee^{-\phi^{2}/b}$ and yielding best fit value $a \simeq 0.035$, $b \simeq 11$. \label{fig:slowroll_2}}
\end{figure}
Fig.\ \ref{fig:slowroll_2} shows the numerical result for the effective inflaton potential $V(\phi)/V_{0}$ together with a fit by a modified Gaussian similar to \eqref{eqn:inflatonpot} with $\phi_0 =0$ for simplicity. It should be noted that the function is plotted with respect to $\phi$ (in units $M_{\text{pl}}=1$), rather than $\tilde{\phi}$. By fitting only a small region of the $\phi$-range around the maximum of the potential, one obtains a best fit value for $b$ (see Fig.\ \ref{fig:slowroll_2}) which is close to the analytic expected value $b=4/\epsilon_2\simeq 18$ for $m=0.1$, while the fit is essentially insensitive to different (small) values of $a$. If the fit range is increased to capture the whole potential, one can see that the shape \eqref{eqn:inflatonpot} still provides an excellent fit of the numerical solution, although the best fit values differ from the analytic ones. This is however to be expected, as \eqref{eqn:inflaton} is valid only in a perturbative regime.

\subsection{Post-inflationary phase}\label{sec:postinflation}
\begin{figure}[htbp]
\centering
\includegraphics[width=.4\textwidth]{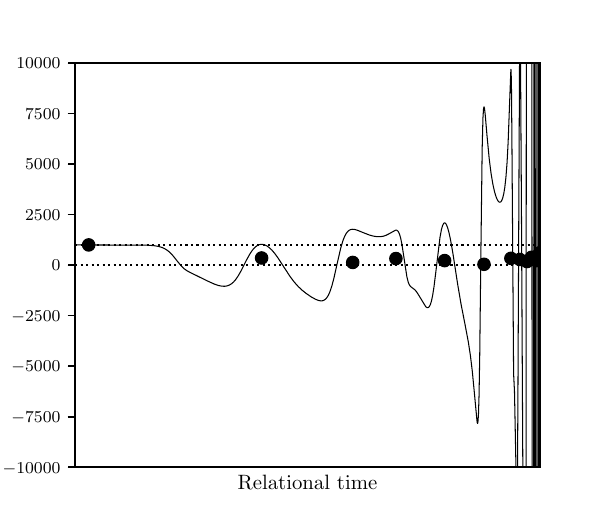}
\\[-4mm]
\includegraphics[width=.4\textwidth]{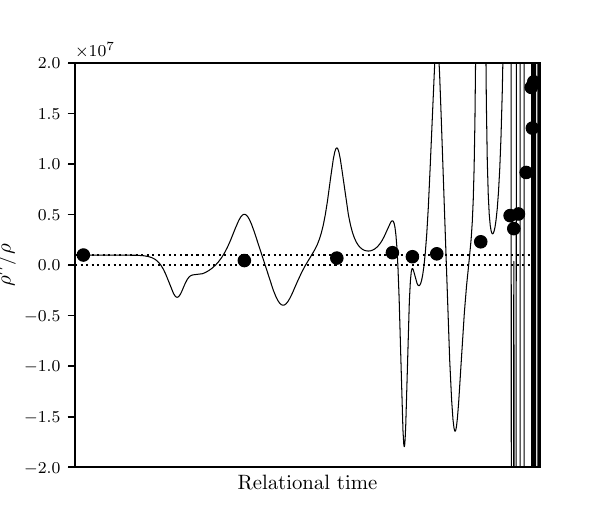}
\caption{$\rho' /\rho$ and $\rho'' /\rho$, with cumulative averages over periods as dots. The dotted line marks $0$. Microscopic parameter and initial conditions are $E=1000.0$, $\vartheta=0.0$, $\lambda=1.0$, $m=1.0$, $l=5.0$, and $\rho_{0}=10.0$, $\rho'_0=E\rho_0$ $\theta_0=\pi/2$, $\theta'_0=1.0$, respectively.\label{fig:oscillation_1}}
\end{figure}

So far, our analysis has focused on the inflationary phase. A key question, however, concerns the subsequent evolution of the universe once inflation ends. In this regime, one expects the quantum gravitational effects responsible for the initial accelerated expansion to have sufficiently dissipated, such that the dynamics become effectively classical and governed by the matter content of the universe—in the present model, a minimally coupled, massless, free scalar field. In this section, we provide tentative analytic arguments supporting the emergence of such a classical phase, complemented by preliminary numerical results. We emphasize, however, that these findings should be regarded only as promising indications, and that a more comprehensive analysis will be required to establish definitive conclusions.
\paragraph*{Qualitative considerations.}
Let us start from equations \eqref{eqn:fundamentalequations}, or equivalently, \eqref{eqn:xdynamics}. A phase of scalar field domination is characterized by the condition $\rho''=\mu^2\rho$, or, equivalently by
\begin{equation}
    f\diff_xy=-2y^2\,,
\end{equation}
where $f\equiv z^2/\rho^2$, and $\diff_x\equiv \diff/\diff x$. One can easily verify that this condition is not dynamically preserved by the system of differential equations in \eqref{eqn:xdynamics}. However, in this regime, the dynamical instability leads to very fast phase changes. Similarly to the arguments used in Sec.\ \ref{sec:stability} and App.\ \ref{app:[proof]}, let us therefore consider a scenario where we average the dynamical equations over a period of the phase, and where the equation above is only valid on average:
\begin{equation}
    \langle f\diff_xy\rangle=-2\langle y^2\rangle\,.
\end{equation}
The resulting averaged equations take the form
\begin{subequations}
\begin{align}
    0&=-\langle y^2\rangle+\langle f^2\rangle\\
    0&=\frac{1}{2}\langle\diff_x(f^2)\rangle+4\langle f y\rangle\,.
\end{align}
\end{subequations}
Let us consider the ansatz $f=a(x) F(x)$, $y=b(x) Y(x)$, where $\phi$ and $\psi$ are slowly varying functions of $x$. Let us define the averages $f_0\equiv \langle F\rangle$, $y_0\equiv \langle Y\rangle$, $k_{fy}\equiv \langle F Y\rangle$, and $k'_{fy}\equiv \langle F\diff_x Y\rangle$. Then the relevant system of equations becomes
\begin{subequations}
\begin{align}
    a^2f_0^2&=b^2y_0^2\\
    f_0\diff_xa&=-4b k_{fy}\\
    k_{fy}a\diff_xb +ab k'_{fy}&=-2b^2 y_0\,.
\end{align}
\end{subequations}
These equations can be immediately solved to obtain
\begin{equation}
    a(x)=a_0 e^{\alpha x}=\sigma b(x)\,,
\end{equation}
where $\sigma=\pm 1$ and $\alpha=-4\sigma k_{fy}/y_0$. This then leads to a consistency condition on the remaining variables of the form
\begin{equation}
    k_{fy}\alpha+k'_{fy}=-2\sigma f_0\,.
\end{equation}
This shows the existence of the required averaged solutions, as long as $\vert \alpha\vert$ is much smaller than the typical phase of oscillations in the trigonometric terms in \eqref{eqn:xdynamics}.
\paragraph*{Numerical considerations.}
Fig.\ \ref{fig:oscillation_1} shows numerical results for $\rho'/\rho$ and $\rho''/\rho$ in the post-inflationary phase. Due to numerical limitations and changing angular frequency, it is very difficult to precisely quantify the averaged dynamics. The attempt shown here uses cumulative averages over ``periods'' identified by the maxima of the data. This, together with numerical instability near the end, limits the reliability of the numerical results in this regime. Keeping this limitations in mind, the above results seem to suggest the above analytic arguments, although the values are inaccurate. Indeed, one can see from Fig.\ \ref{fig:oscillation_1} that the averaged $\rho'/\rho$ and $\rho''/\rho$ are roughly constant and satisfy $(\rho'/\rho)^2\sim \rho''/\rho$, as expected. More precisely, averaging over the first few data points yields $\rho'/\rho \simeq 0.39,E$ and $\rho''/\rho \simeq 0.89,E^{2}$, where $E$ is a constant, so the two quantities are not exactly equal but of comparable magnitude. Given the sensitivity of these estimates to the specific averaging prescription, we remark again that these results should only be considered as qualitative and preliminary indications of the manifestation of the expected cosmological behavior.

\section{Discussion and conclusions}
In this work, we have investigated the emergence of cosmic acceleration from interacting mean-field GFT models generalizing previously studied pseudosimplicial (PS) and pseudotensorial (PT) cases. The qualitative behavior of these models depends on two key parameters: the interaction order 
$l$ and the angular phase integer $m$. We have shown that for $l=5$, the (relational) dynamics with respect to a minimally coupled massless scalar field admit fixed points corresponding to a cosmological dS phase. However, this phase acts as an attractor only when $m\le 0$.

For these values of $m$, the universe becomes effectively dominated at late times by an emergent dynamical dark-energy component, which approaches a cosmological constant in the asymptotic future. The detailed dynamical features depend on the sign of $1+3m/4$. The resulting equation-of-state parameter $w(z)$ exhibits a distinct suppression factor at late times, modulated respectively by oscillatory-logarithmic functions, purely logarithmic contributions, or mildly increasing monomials, depending on whether $1+3m/4$ is negative, zero, or positive. For the large region $2(3-2\sqrt{6})/3\le m<0$ and $m\le -6$ (thus including the PS case), the late times dark energy dynamics exhibits phantom-like behavior. We can conclude that the emergence of a viable late-time cosmology with an accelerated phase driven purely by quantum gravity effects is a rather general outcome of mean-field GFT dynamics, even once a non-trivial condensate phase dynamics is included. Indeed, it is precisely this non-trivial phase evolution that yields a broad class of phantom-like dark energy behaviors at late times, already for a single representation mode, thereby extending the two-mode mechanism identified in previous work \cite{Oriti:2021rvm,Pang:2025jtk}. This novel quantum gravity mechanism for (phantom) late-time acceleration thus appears remarkably robust.

When $m>0$, the dS phase becomes unstable and the system naturally exits the accelerating regime. We find that for $m\ll 1$, this leads to a quasi-dS epoch consistent with an emergent slow-roll inflationary dynamics, characterized by a constant second slow-roll parameter $\epsilon_2=6(\mu-1)$. This inflationary regime can be effectively reformulated as single-field slow-roll inflation driven by an emergent inflaton with a Gaussian-type potential whose amplitude includes a quadratic prefactor. Interestingly, the inflaton field corresponds directly to the phase degree of freedom of the GFT mean field, reinforcing its interpretation as a collective excitation of the underlying quantum-geometric structure. Finally, we provide preliminary evidence that, once inflation ends, the Universe transitions into a regime dominated by the minimally coupled scalar field serving as the relational clock, governed by a standard expanding Friedmann dynamics, as expected.

All the above results are supported by both perturbative analytic studies and non-perturbative numerical analyses. The latter, however, become significantly more challenging in the post-inflationary regime, where the dynamical system exhibits rapid and highly non-linear transitions. A detailed investigation of this regime will be pursued in future work.

Despite this, the predictions for the emergent dynamical dark-energy sector are already sufficiently robust to allow for comparison with current DESI data, and potentially with future Euclid observations. In principle, the emergent slow-roll phase could also be tested against CMB data, provided one treats the emergent inflaton as an effective quantum field in close analogy with standard inflationary cosmology. It is crucial, however, to further validate this effective description by deriving additional evidence directly within the full quantum-gravity framework, following the initial steps taken in \cite{Jercher:2023kfr,Jercher:2023nxa}. This will require a number of formal improvements of the underlying TGFT framework, especially to gain more control over models including realistic matter content.

Empirical constraints from these observational tests would translate into meaningful bounds on the parameters $m$ and $l$, which directly characterize the underlying quantum-geometric dynamics, as well as on $z_q$, which encodes the present density of quantum gravity atoms and therefore provides direct information about genuinely quantum-gravitational features of our universe. This would enable—for the first time—a quantitatively data-driven approach to quantum gravity model building. Moreover, these results offer a new conceptual understanding of the mechanisms behind cosmic acceleration, thus opening to better physically motivated inflationary and dark energy models.  In this sense, these findings represent an important concrete step towards connecting full quantum gravity with cosmological observations.

\label{sec:conclusions}
\acknowledgments{
We thank Edward Wilson-Ewing and Elisa Ferreira for helpful discussions. L.M.~acknowledges support from the Kavli Insitute for the Physics and Mathematics of the Universe, from the Okinawa Institute of Science and Technology Graduate University, and from the John Templeton Foundation, through ID\# 62312 grant as part of the \href{https://www.templeton.org/grant/the-quantum-information-structure-of-spacetime-qiss-second-phase}{\textit{`The Quantum Information Structure of Spacetime'} Project (QISS)}.}
The code used to perform the analyses in this work is available at \url{https://github.com/tomrobe/acceleration}.

\appendix
\section{The emergent cosmology dictionary}\label{app:dictionary}
In this appendix, we provide the main identities to convert equations in terms of quantum gravity data (i.e.\ GFT mean-field wavefunction) to cosmological quantities, within a homogeneous and isotropic setting. As explained in Sec.\ \ref{sec:interactinggft}, the main identity is given by
\begin{equation}
    V_0a^3=\mathfrak{v}\rho^2\,,
\end{equation}
where $V_0$ is a fiducial cosmological volume, $\mathfrak{v}_\upsilon$ are eigenvalues of the quantum volume operator on an isotropic volume element (tetrahedron), and where we are working within a regime where a single model (labeled by $\upsilon_o$, which we dropped in the above equation). The above expression determines $a(\rho)$, and thus also allows us to express quantities such as the number of $e$-folds $\Delta N$ and the redshift $z$ as functions of $\rho$:
\begin{subequations}\label{eqn:nandz}
    \begin{align}
        \Delta N&=\log\frac{a_f}{a_i}=\frac{2}{3}\log\frac{\rho_f}{\rho_i}\\
        1+z&=\frac{a_0}{a}=\left(\frac{\rho_0}{\rho}\right)^{2/3}\,.
    \end{align}
\end{subequations}
Equivalently, one can invert the above relation to write
\begin{equation}\label{eqn:rhotoz}
    \rho^2=\left(\frac{1+z_q}{1+z}\right)^3\,,\qquad z_q\equiv z(\rho=1)\,.
\end{equation}
On the quantum gravity side, dynamics is described relationally, with respect to a minimally coupled massless scalar field clock $\chi$. On the other hand, quantities in cosmology are often expressed with respect to cosmic time $t$. We denote derivatives with respect to the clock $\chi$ with a $'\equiv \diff/\diff\chi$, while derivatives with respect to $t$ with a $\dot{}\equiv \diff/\diff t$. For instance, the relational Hubble parameter $\mathcal{H}\equiv V'/(3V)$ is related to the cosmic one $H=\dot{V}/3V$ by $H=\dot{\chi} \mathcal{H}$, and thus
\begin{equation}\label{eqn:friedmann}
    H^2=\dot{\chi}^2\mathcal{H}^2=\frac{4}{9}\dot{\chi}^2\left(\frac{\rho'}{\rho}\right)^2=\frac{4\pi_\chi^2}{9\mathfrak{v}^2}\left(\frac{\rho'}{\rho^3}\right)^2\,,
\end{equation}
since $\dot{\chi}V=\text{const}\equiv\pi_\chi$. Note that as a consequence of this, we also have $\ddot{\chi}=-3\dot{\chi}^2\mathcal{H}$. This can be used to obtain $\dot{H}=\dot{\chi}^2(\mathcal{H}'-3\mathcal{H}^2)$, which in turn leads to \cite{Martin:2013tda}
\begin{equation}
    \epsilon_1\equiv \epsilon_H\equiv -\frac{\dot H}{H^2}=\frac{3}{2}\left[3-\frac{\rho''\rho}{(\rho')^2}\right]\,.
\end{equation}
From this equation one can immediately obtain the effective equation of state parameter,
\begin{equation}\label{eqn:eosparameter}
    w\equiv\frac{2}{3}\epsilon_1-1=2-\frac{\rho''\rho}{(\rho')^2}\,.
\end{equation}
Therefore, a cosmological state with constant $w$ corresponds to a mean-field dynamics characterized (in an appropriate limit) by 
\begin{equation}\label{eq:eosde}
    \frac{\rho''}{\rho} = \left(w - 2\right) \left(\frac{\rho'}{\rho}\right)^{2}\quad\leftrightarrow\quad (\rho'\rho^{2-w})'=0
\end{equation}
Finally, as shown in \cite{Mukhanov:2013tua}, given $\epsilon_1$ (or equivalently $w$), one can define a scalar field $\phi$ with potential $V(\phi)$ driving the associated dynamics:
\begin{subequations}\label{eqn:emergentscalar}
\begin{align}
    \phi(N)-\phi_0&=\pm \sqrt{3}M_{\text{pl}}\int^N\diff N'\sqrt{2\epsilon_1(N')/3}\,,\\
    \frac{V(N)}{V_0}&=\left[1-\frac{\epsilon_1(N)}{3}\right]e^{-2\int^N\diff N'\epsilon_1(N')}\,.
\end{align}
\end{subequations}
\section{Proof of weak attraction}\label{app:[proof]}
To prove that the fixed points \eqref{eqn:fixedpoints} are (local) weak forward attractors for the system with $1+3m/4<0$, it is convenient to define the two independent solutions in \eqref{eqn:deltax} as $\xi_{1,2}$, so that $\xi\equiv A \xi_1+B\xi_2$, and we can write $ U(\chi,\chi_0)=X(\chi)X^{-1}(\chi_0)$, where
\begin{equation}
    X(\chi)\equiv (X_1(\chi),X_2(\chi))\,,\qquad X_i(\chi)\equiv \begin{pmatrix}
        \xi_i(\chi)\\
        \xi_i'(\chi)
    \end{pmatrix}\,,
\end{equation}
for $i=1,2$. For $1+3m/4<0$, the components of $U$ have the form $a(\rho)e^{\pm i\Phi(\rho)}$, where $a(\rho)$ is either  $\mathcal{O}(\rho^{-2})$ or $\mathcal{O}(1)$ at late times. First, we show that $U(\chi,\chi_0)\rightharpoonup 0$ weakly, i.e., that for any $\ell\in\mathcal{X}^*$ (where we recall that $\mathcal{X}$ is the local space of $(\xi,\zeta)$ data around a fixed point), $\ell(U(\chi,\chi_0)v)\to 0$ for fixed $v\in\mathcal{X}$. Since in finite dimension any linear functional takes the form $\ell_w(v)\equiv \langle w,v\rangle$ for some $w$, we need to compute $\langle w,U(\chi,\chi_0)v\rangle$. To do this, we consider the following identity:
\begin{equation}
    \langle w,U(\chi,\chi_0)v\rangle=
    \lim_{\epsilon\to 0}\int_{\mathbb{R}}\diff\chi'm_{\chi,\epsilon}(\chi')\langle w,U(\chi',\chi_0)v\rangle\,,
\end{equation}
where $m_{t,\epsilon}$ is a mollifier supported in $[t-\epsilon,t+\epsilon]$ with unit integral, and with $\epsilon\Phi'\to \infty$ as $\rho\to\infty$. The right-hand-side physically represents a $\chi$-local measurement with finite (but small) uncertainty $2\epsilon$. The above integral is a finite sum of integrals of the form
\begin{equation}
    I(\chi)=\int_\mathbb{R}\diff\chi' m_{\chi,\epsilon}(\chi')a(\chi')e^{i\Phi(\chi)\Phi_{\text{rel}}(\chi')}\,,
\end{equation}
where $\Phi_{\text{rel}}(\chi')=\Phi(\chi')/\Phi(\chi)$ is the relative phase around $\chi$, multiplied by the large parameter $\Phi(\chi)$. The function $m_{\chi,\epsilon}(\chi')a(\chi')$ is a smooth function with compact support, while $\Phi_{\text{rel}}(\chi')$ is a smooth and non-stationary function on the support of $m_{\chi,\epsilon}$. One can thus apply the Riemann-Lebesgues lemma (non-stationary phase approximation) to conclude that, since $\Phi(\chi)\to\infty$ as $\rho\to\infty$, in the same limit $I(\chi)\to 0$. As a consequence, we also have that 
\begin{equation}
    \langle w,U(\chi,\chi_0)v\rangle\to 0\,,\qquad\rho\to\infty\,,
\end{equation}
As in finite dimensions pointwise convergence of matrix elements implies weak operator convergence, we conclude that $U(\chi,\chi_0)\rightharpoonup 0$.

Now, let $B$ a bounded set of initial data, say $\Vert v\Vert\le R$ for any $v\in B$. For any linear functional $\ell_w(v)=\langle w,v\rangle$, then
\begin{equation}\label{eqn:boundnorm}
    \mathrm{sup}_{v\in B}\vert\ell_w(U(\chi,\chi_0)v)\vert\le R\Vert U(\chi,\chi_0)^*w\Vert\,.
\end{equation}
But from the above result $U(\chi,\chi_0)^\star w\to 0$ componentwise, and in finite dimensions this implies $\Vert  U(\chi,\chi_0)^*w\Vert\to 0$. Now, recall that the weak distance between a point $\{0\}$ and a set $S\in\mathcal{X}$ is defined by
\begin{equation}\label{eqn:weakdistance}
    \mathrm{dist}_{\mathrm{w}}(S,\{0\})=\mathrm{sup}_{\ell\in\mathcal{X}^*,\Vert \ell \Vert \le 1}\mathrm{sup}_{x\in S}| \ell(x)|\,.
\end{equation}
A weak forward attractor is a fixed point characterized by weakly vanishing perturbations for any bounded set $B\in \mathcal{X}$:
\begin{equation}
    \mathrm{dist}_{\mathrm{w}}(U(\chi,\chi_0)B,\{0\})\to 0\,,\quad\rho\to\infty\,.
\end{equation}
Using equations \eqref{eqn:boundnorm} and \eqref{eqn:weakdistance}, it is then clear that the fixed point \eqref{eqn:fixedpoints} is a weak forward attractor for $1+3m/4<0$.
\section{Perturbative computations}\label{sec:computations}
From equation \eqref{eqn:firstordery}, we see that linear perturbations in $y$ are suppressed at late times. Hence, here we will look for corrections in $y$ at second order, triggered by perturbations of the form \eqref{eqn:deltax} and \eqref{eqn:deltaz}. As the computations differ slightly for the cases $1+3m\ge 0$ (which includes the PS and PT cases of Sec.\ \ref{sec:interactinggft}) and $1+3m<0$, we consider them separately.
\paragraph*{$1+3m/4<0$.}
In this case, we make the following ansatz for $y$:
\begin{equation}\label{eqn:yofg}
    y=\bar{y}(1+\rho^{-4}g(\Phi))\,,\qquad\vert g(\Phi)\vert\ll 1\,.
\end{equation}
where $\Phi=\beta\log\rho$. Expanding equation \eqref{eqn:yprime} at second order in $\xi$ and keeping only first order contributions in $g$ (equivalently, assuming $g\sim \mathcal{O}((\rho^2\xi)^2)\sim \mathcal{O}(\zeta^2)$), and plugging in the above ansatz together with equations \eqref{eqn:deltax} and \eqref{eqn:deltaz} for $1+3m/4<0$, we obtain
\begin{equation}
    2g(\Phi)+\beta\frac{\diff}{\diff\Phi}g(\Phi)+s(\Phi)=0
\end{equation}
where $s(\Phi)=\kappa_A\cos^2\Phi+\kappa_B\sin^2\Phi+\kappa_{AB}\sin\Phi\cos\Phi$, with $\kappa_A\equiv 3m^2A^2/2-4c_A^2$, $\kappa_B\equiv 3m^2B^2/2-4c_B^2$, and $\kappa_{AB}\equiv 3m^2AB-8c_Ac_B$.
The homogeneous solution is exponentially suppressed, so we focus on the non-homogeneous solution, which is given by
\begin{align}\label{eqn:g}
    g(\Phi)&=-\frac{\kappa_A+\kappa_B}{4}+\frac{(\kappa_B+\beta\kappa_{AB}-\kappa_A)\cos2\Phi}{4(1+\beta^2)}\nonumber\\
    &\quad+
    \frac{(\beta\kappa_B-\beta \kappa_A-\kappa_{AB})\sin2\Phi}{4(1+\beta^2)}\,.
\end{align}
Thus, $\vert y-\bar{y}\vert/\bar{y}\sim\mathcal{O}(\rho^{-4})$, which is asymptotically dominating over first order corrections of the form \eqref{eqn:firstordery}. From the above equation, we obtain
\begin{equation}\label{eqn:yprimecase3}
    \frac{y'}{\rho^2\bar{y}^2}=\rho^{-4}\left[(\kappa_A+\kappa_B)+\tilde{\kappa}_1\cos2\Phi+\tilde{\kappa}_2\sin2\Phi\right]\,,
\end{equation}
where
\begin{align*}
    \tilde{\kappa}_1&\equiv [2(1+\beta^2)]^{-1}[(2-\beta^2)(\kappa_A-\kappa_B)-3\beta\kappa_{AB}]\,,\\
    \tilde{\kappa}_2&\equiv[2(1+\beta^2)]^{-1}[3\beta(\kappa_A-\kappa_B)+(2-\beta^2)\kappa_{AB}]\,.
\end{align*}
We thus conclude that, at late times, $y$ and $y'/(\rho^2y^2)$ tend to $\bar{y}$ and $0$, respectively, confirming the attractor nature of the emergent dS phase. The above equations can be used to explicitly compute cosmological parameters. For instance, they allow us to compute the parameters entering the equation of state parameter for the emergent dark energy:
\begin{subequations}\label{eqn:deltawk}
    \begin{align}
        \delta w_0&=-(\kappa_A+\kappa_B)\,,\\
        \delta w_1&=-\tilde{\kappa}_1\,,\\
        \delta w_1&=-\tilde{\kappa}_2\,.
    \end{align}
\end{subequations}
\paragraph*{$1+3m\ge/4 0$.} The analysis in this case is analogous to the one above, although computations can be significantly simplified by defining $f\equiv z/\rho^2$, and rewriting equations \eqref{eqn:fundamentalequations} as
\begin{align}\label{eqn:xdynamics}
    f\diff_xy&=-3y^2+f^2+\lambda\cos(\vartheta+ m x)\,,\\
    f\diff_xf&=-4fy+\lambda\sin(\vartheta+ m x)\,,
\end{align}
where we have assumed $f\neq 0$, and defined $\diff_x\equiv \diff/\diff x$. As we did before, we assume $x=\bar{x}_n+\xi$, with $\vert\xi\vert\ll 1$ so that we can expand the above equations as
\begin{subequations}\label{eqn:xdynamicslinearized}
\begin{align}
    f\diff_\xi y&=-3y^2+f^2+3\bar{y}^2(1-(m\xi)^2/2)\\
    f\diff_\xi f&=-4fy+3\bar{y}^2 m \xi\,.
\end{align}
\end{subequations}
To find unstable solutions to the above equation, we can note from equations \eqref{eqn:deltax} and \eqref{eqn:deltaz} that the unstable branches of the perturbative solutions for $1+3m/4\ge 0$ are characterized (in terms of the variables $(f,\xi)$) by
\begin{subequations}\label{eqn:unstableansatz}
\begin{equation}\label{eqn:fofxi}
    f(\xi)=\rho^2\zeta=-2\bar{y}(1-\mu)\xi\,,
\end{equation} 
As in the $1+3m/4<0$ case, we expect second-order corrections to $y$. Thus, we make the following ansatz: 
\begin{equation}\label{eqn:yofxi}
    y(\xi)=\bar{{y}}(1+y_1m^2\xi^2)\,.
\end{equation}
\end{subequations}
Plugging equations \eqref{eqn:unstableansatz} into \eqref{eqn:xdynamicslinearized} and using equations \eqref{eqn:fixedpoints} to simplify, we obtain, at lowest order,
\begin{equation}\label{eqn:y1}
    y_1=-2\left[\frac{3}{8}-\frac{(1-\mu)^2}{m^2}\right](1+2\mu)^{-1}\,.
\end{equation}
Using equations \eqref{eqn:deltax} for $1+3m/4\ge 0$, we see that only for $\mu-1< 0$ (equivalently, $m<0)$, $y\to \bar{y}$ asymptotically and $\bar{y}$ is an attractor. Moreover, we can compute
\begin{equation}\label{eqn:yprimeothercases}
    \frac{y'}{\rho^2y^2}=\frac{f\diff_\xi y}{y^2}=-2y_1m^2(1-\mu)\xi^2\,,
\end{equation}
which tends to zero asymptotically for $\mu-1<0$.

\bibliographystyle{JHEP.bst}
\bibliography{biblio}

@book{Dodelson:2020bqr,
    author = "Dodelson, Scott and Schmidt, Fabian",
    title = "{Modern Cosmology}",
    doi = "10.1016/C2017-0-01943-2",
    publisher = "Academic Press",
    year = "2020"
}

@article{Starobinsky:1980te,
    author = "Starobinsky, Alexei A.",
    editor = "Khalatnikov, I. M. and Mineev, V. P.",
    title = "{A New Type of Isotropic Cosmological Models Without Singularity}",
    doi = "10.1016/0370-2693(80)90670-X",
    journal = "Phys. Lett. B",
    volume = "91",
    pages = "99--102",
    year = "1980"
}

@article{LINDE1982389,
title = {A new inflationary universe scenario: A possible solution of the horizon, flatness, homogeneity, isotropy and primordial monopole problems},
journal = {Physics Letters B},
volume = {108},
number = {6},
pages = {389-393},
year = {1982},
issn = {0370-2693},
doi = {https://doi.org/10.1016/0370-2693(82)91219-9},
url = {https://www.sciencedirect.com/science/article/pii/0370269382912199},
author = {A.D. Linde}
}

@article{guth,
  title = {Inflationary universe: A possible solution to the horizon and flatness problems},
  author = {Guth, Alan H.},
  journal = {Phys. Rev. D},
  volume = {23},
  issue = {2},
  pages = {347--356},
  numpages = {0},
  year = {1981},
  month = {Jan},
  publisher = {American Physical Society},
  doi = {10.1103/PhysRevD.23.347},
  url = {https://link.aps.org/doi/10.1103/PhysRevD.23.347}
}

@article{Riotto:2002yw,
    author = "Riotto, Antonio",
    editor = "Dvali, G. and Perez-Lorenzana, Abdel and Senjanovic, G. and Thompson, G. and Vissani, F.",
    title = "{Inflation and the theory of cosmological perturbations}",
    eprint = "hep-ph/0210162",
    archivePrefix = "arXiv",
    reportNumber = "DFPD-TH-02-22",
    journal = "ICTP Lect. Notes Ser.",
    volume = "14",
    pages = "317--413",
    year = "2003"
}

@article{Martin:2000xs,
    author = "Martin, Jerome and Brandenberger, Robert H.",
    title = "{The TransPlanckian problem of inflationary cosmology}",
    eprint = "hep-th/0005209",
    archivePrefix = "arXiv",
    doi = "10.1103/PhysRevD.63.123501",
    journal = "Phys. Rev. D",
    volume = "63",
    pages = "123501",
    year = "2001"
}

@article{Martin:2012bt,
    author = "Martin, Jerome",
    title = "{Everything You Always Wanted To Know About The Cosmological Constant Problem (But Were Afraid To Ask)}",
    eprint = "1205.3365",
    archivePrefix = "arXiv",
    primaryClass = "astro-ph.CO",
    doi = "10.1016/j.crhy.2012.04.008",
    journal = "Comptes Rendus Physique",
    volume = "13",
    pages = "566--665",
    year = "2012"
}

@article{DiValentino:2021izs,
    author = {E. Di Valentino and O. Mena and S. Pan and L. Visinelli and W. Yang and A. Melchiorri and D. F. Mota and A. G. Riess and J. Silk},
    title = {In the realm of the Hubble tension—a review of solutions},
    journal = {Class. Quant. Grav.},
    volume = 38,
    number = 15,
    pages = 153001,
    year = 2021,
    doi = {10.1088/1361-6382/ac086d},
    eprint = {2103.01183},
    archivePrefix = {arXiv},
    primaryClass = {astro-ph.CO}
}

@article{DESI:2025zgx,
    author = "Abdul Karim, M. and others",
    collaboration = "DESI",
    title = "{DESI DR2 results. II. Measurements of baryon acoustic oscillations and cosmological constraints}",
    eprint = "2503.14738",
    archivePrefix = "arXiv",
    primaryClass = "astro-ph.CO",
    reportNumber = "FERMILAB-PUB-25-0169-PPD",
    doi = "10.1103/tr6y-kpc6",
    journal = "Phys. Rev. D",
    volume = "112",
    number = "8",
    pages = "083515",
    year = "2025"
}

@article{Avsajanishvili:2023jcl,
    author = "Avsajanishvili, Olga and Chitov, Gennady Y. and Kahniashvili, Tina and Mandal, Sayan and Samushia, Lado",
    title = "{Observational Constraints on Dynamical Dark Energy Models}",
    eprint = "2310.16911",
    archivePrefix = "arXiv",
    primaryClass = "astro-ph.CO",
    doi = "10.3390/universe10030122",
    journal = "Universe",
    volume = "10",
    number = "3",
    pages = "122",
    year = "2024"
}

@article{Planck:2018jri,
    author = "Akrami, Y. and others",
    collaboration = "Planck",
    title = "{Planck 2018 results. X. Constraints on inflation}",
    eprint = "1807.06211",
    archivePrefix = "arXiv",
    primaryClass = "astro-ph.CO",
    doi = "10.1051/0004-6361/201833887",
    journal = "Astron. Astrophys.",
    volume = "641",
    pages = "A10",
    year = "2020"
}

@article{ACT:2025tim,
    author = "Calabrese, Erminia and others",
    collaboration = "ACT",
    title = "{The Atacama Cosmology Telescope: DR6 Constraints on Extended Cosmological Models}",
    eprint = "2503.14454",
    archivePrefix = "arXiv",
    primaryClass = "astro-ph.CO",
    reportNumber = "FERMILAB-PUB-25-0157-PPD",
    month = "3",
    year = "2025"
}

@article{deCesare:2016rsf,
    author = "de Cesare, Marco and Pithis, Andreas G. A. and Sakellariadou, Mairi",
    title = "{Cosmological implications of interacting Group Field Theory models: cyclic Universe and accelerated expansion}",
    eprint = "1606.00352",
    archivePrefix = "arXiv",
    primaryClass = "gr-qc",
    reportNumber = "KCL-PH-TH-2016-32",
    doi = "10.1103/PhysRevD.94.064051",
    journal = "Phys. Rev. D",
    volume = "94",
    number = "6",
    pages = "064051",
    year = "2016"
}

@article{Oriti:2021rvm,
    author = "Oriti, Daniele and Pang, Xiankai",
    title = "{Phantom-like dark energy from quantum gravity}",
    eprint = "2105.03751",
    archivePrefix = "arXiv",
    primaryClass = "gr-qc",
    doi = "10.1088/1475-7516/2021/12/040",
    journal = "JCAP",
    volume = "12",
    number = "12",
    pages = "040",
    year = "2021"
}

@article{Pang:2025jtk,
    author = "Pang, Xiankai and Oriti, Daniele",
    title = "{Late-time cosmic acceleration from quantum gravity}",
    eprint = "2502.12419",
    archivePrefix = "arXiv",
    primaryClass = "gr-qc",
    doi = "10.1088/1361-6382/adecdb",
    journal = "Class. Quant. Grav.",
    volume = "42",
    number = "15",
    pages = "155003",
    year = "2025"
}

@article{Martin:2013tda,
    author = "Martin, Jerome and Ringeval, Christophe and Vennin, Vincent",
    title = "{Encyclop\ae{}dia Inflationaris}",
    eprint = "1303.3787",
    archivePrefix = "arXiv",
    primaryClass = "astro-ph.CO",
    doi = "10.1016/j.dark.2014.01.003",
    journal = "Phys. Dark Univ.",
    volume = "5-6",
    pages = "75--235",
    year = "2014"
}

@article{Heisenberg:2022gqk,
    author = "Heisenberg, Lavinia and Villarrubia-Rojo, Hector and Zosso, Jann",
    title = "{Can late-time extensions solve the H0 and {\ensuremath{\sigma}}8 tensions?}",
    eprint = "2202.01202",
    archivePrefix = "arXiv",
    primaryClass = "astro-ph.CO",
    doi = "10.1103/PhysRevD.106.043503",
    journal = "Phys. Rev. D",
    volume = "106",
    number = "4",
    pages = "043503",
    year = "2022"
}

@article{Ketov:2014qha,
    author = "Ketov, Sergei V. and Terada, Takahiro",
    title = "{Inflation in supergravity with a single chiral superfield}",
    eprint = "1406.0252",
    archivePrefix = "arXiv",
    primaryClass = "hep-th",
    reportNumber = "IPMU14-0128, UT-14-27",
    doi = "10.1016/j.physletb.2014.07.036",
    journal = "Phys. Lett. B",
    volume = "736",
    pages = "272--277",
    year = "2014"
}

@article{Ellis:2014rxa,
    author = "Ellis, John and Garc{\'\i}a, Marcos A. G. and Nanopoulos, Dimitri V. and Olive, Keith A.",
    title = "{Resurrecting Quadratic Inflation in No-Scale Supergravity in Light of BICEP2}",
    eprint = "1403.7518",
    archivePrefix = "arXiv",
    primaryClass = "hep-ph",
    reportNumber = "KCL-PH-TH-2014-11, LCTS-2014-11, CERN-PH-TH-2014-049, ACT-3-14, FTPI-MINN-14-10, UMN-TH-3332-14",
    doi = "10.1088/1475-7516/2014/05/037",
    journal = "JCAP",
    volume = "05",
    pages = "037",
    year = "2014"
}

@article{Marchetti:2020xvf,
    author = {Marchetti, Luca and Oriti, Daniele and Pithis, Andreas G. A. and Th\"urigen, Johannes},
    title = "{Phase transitions in tensorial group field theories: Landau-Ginzburg analysis of models with both local and non-local degrees of freedom}",
    eprint = "2110.15336",
    archivePrefix = "arXiv",
    primaryClass = "gr-qc",
    doi = "10.1007/JHEP12(2021)201",
    journal = "JHEP",
    volume = "21",
    pages = "201",
    year = "2020"
}

@article{Oriti:2016qtz,
    author = "Oriti, Daniele and Sindoni, Lorenzo and Wilson-Ewing, Edward",
    title = "{Emergent Friedmann dynamics with a quantum bounce from quantum gravity condensates}",
    eprint = "1602.05881",
    archivePrefix = "arXiv",
    primaryClass = "gr-qc",
    doi = "10.1088/0264-9381/33/22/224001",
    journal = "Class. Quant. Grav.",
    volume = "33",
    number = "22",
    pages = "224001",
    year = "2016"
}

@article{Marchetti:2022nrf,
    author = {Marchetti, Luca and Oriti, Daniele and Pithis, Andreas G. A. and Th\"urigen, Johannes},
    title = "{Mean-Field Phase Transitions in Tensorial Group Field Theory Quantum Gravity}",
    eprint = "2211.12768",
    archivePrefix = "arXiv",
    primaryClass = "gr-qc",
    doi = "10.1103/PhysRevLett.130.141501",
    journal = "Phys. Rev. Lett.",
    volume = "130",
    number = "14",
    pages = "141501",
    year = "2023"
}

@article{Gielen:2013naa,
    author = "Gielen, Steffen and Oriti, Daniele and Sindoni, Lorenzo",
    title = "{Homogeneous cosmologies as group field theory condensates}",
    eprint = "1311.1238",
    archivePrefix = "arXiv",
    primaryClass = "gr-qc",
    reportNumber = "AEI-2013-259",
    doi = "10.1007/JHEP06(2014)013",
    journal = "JHEP",
    volume = "06",
    pages = "013",
    year = "2014"
}

@article{Perez:2003vx,
    author = "Perez, Alejandro",
    title = "{Spin foam models for quantum gravity}",
    eprint = "gr-qc/0301113",
    archivePrefix = "arXiv",
    doi = "10.1088/0264-9381/20/6/202",
    journal = "Class. Quant. Grav.",
    volume = "20",
    pages = "R43",
    year = "2003"
}

@article{Freidel:2005qe,
    author = "Freidel, Laurent",
    title = "{Group field theory: An Overview}",
    eprint = "hep-th/0505016",
    archivePrefix = "arXiv",
    doi = "10.1007/s10773-005-8894-1",
    journal = "Int. J. Theor. Phys.",
    volume = "44",
    pages = "1769--1783",
    year = "2005"
}

@inproceedings{Oriti:2011jm,
    author = "Oriti, Daniele",
    title = "{The microscopic dynamics of quantum space as a group field theory}",
    booktitle = "{Foundations of Space and Time: Reflections on Quantum Gravity}",
    eprint = "1110.5606",
    archivePrefix = "arXiv",
    primaryClass = "hep-th",
    pages = "257--320",
    month = "10",
    year = "2011"
}

@article{Oriti:2013aqa,
    author = "Oriti, Daniele",
    title = "{Group field theory as the 2nd quantization of Loop Quantum Gravity}",
    eprint = "1310.7786",
    archivePrefix = "arXiv",
    primaryClass = "gr-qc",
    reportNumber = "AEI-2013-256",
    doi = "10.1088/0264-9381/33/8/085005",
    journal = "Class. Quant. Grav.",
    volume = "33",
    number = "8",
    pages = "085005",
    year = "2016"
}

@article{Perez:2012wv,
    author = "Perez, Alejandro",
    title = "{The Spin Foam Approach to Quantum Gravity}",
    eprint = "1205.2019",
    archivePrefix = "arXiv",
    primaryClass = "gr-qc",
    doi = "10.12942/lrr-2013-3",
    journal = "Living Rev. Rel.",
    volume = "16",
    pages = "3",
    year = "2013"
}

@article{Pithis:2019tvp,
    author = "Pithis, A. G. A. and Sakellariadou, M.",
    title = "{Group field theory condensate cosmology: An appetizer}",
    eprint = "1904.00598",
    archivePrefix = "arXiv",
    primaryClass = "gr-qc",
    reportNumber = "KCL-PH-TH/2019-41",
    doi = "10.3390/universe5060147",
    journal = "Universe",
    volume = "5",
    number = "6",
    pages = "147",
    year = "2019"
}

@article{Oriti:2006se,
    author = "Oriti, Daniele",
    title = "{The Group field theory approach to quantum gravity}",
    eprint = "gr-qc/0607032",
    archivePrefix = "arXiv",
    reportNumber = "DAMTP-2006-54",
    pages = "310--331",
    month = "7",
    year = "2006"
}

@article{Ashtekar:2004eh,
    author = "Ashtekar, Abhay and Lewandowski, Jerzy",
    title = "{Background independent quantum gravity: A Status report}",
    eprint = "gr-qc/0404018",
    archivePrefix = "arXiv",
    doi = "10.1088/0264-9381/21/15/R01",
    journal = "Class. Quant. Grav.",
    volume = "21",
    pages = "R53",
    year = "2004"
}

@phdthesis{Carrozza:2013oiy,
    author = "Carrozza, Sylvain",
    title = "{Tensorial methods and renormalization in Group Field Theories}",
    eprint = "1310.3736",
    archivePrefix = "arXiv",
    primaryClass = "hep-th",
    reportNumber = "tel-00872784, 2013PA112147",
    doi = "10.1007/978-3-319-05867-2",
    school = "Orsay, LPT",
    year = "2013"
}

@article{Baratin:2011tx,
    author = "Baratin, Aristide and Oriti, Daniele",
    title = "{Quantum simplicial geometry in the group field theory formalism: reconsidering the Barrett-Crane model}",
    eprint = "1108.1178",
    archivePrefix = "arXiv",
    primaryClass = "gr-qc",
    reportNumber = "LPT-ORSAY-11-121",
    doi = "10.1088/1367-2630/13/12/125011",
    journal = "New J. Phys.",
    volume = "13",
    pages = "125011",
    year = "2011"
}

@article{Baratin:2010wi,
    author = "Baratin, Aristide and Oriti, Daniele",
    title = "{Group field theory with non-commutative metric variables}",
    eprint = "1002.4723",
    archivePrefix = "arXiv",
    primaryClass = "hep-th",
    doi = "10.1103/PhysRevLett.105.221302",
    journal = "Phys. Rev. Lett.",
    volume = "105",
    pages = "221302",
    year = "2010"
}

@article{Bonzom:2009hw,
    author = "Bonzom, Valentin",
    title = "{Spin foam models for quantum gravity from lattice path integrals}",
    eprint = "0905.1501",
    archivePrefix = "arXiv",
    primaryClass = "gr-qc",
    doi = "10.1103/PhysRevD.80.064028",
    journal = "Phys. Rev. D",
    volume = "80",
    pages = "064028",
    year = "2009"
}

@article{Finocchiaro:2018hks,
    author = "Finocchiaro, Marco and Oriti, Daniele",
    title = "{Spin foam models and the Duflo map}",
    eprint = "1812.03550",
    archivePrefix = "arXiv",
    primaryClass = "gr-qc",
    doi = "10.1088/1361-6382/ab58da",
    journal = "Class. Quant. Grav.",
    volume = "37",
    number = "1",
    pages = "015010",
    year = "2020"
}

@article{Oriti:2014yla,
    author = {Oriti, Daniele and Ryan, James P. and Th\"urigen, Johannes},
    title = "{Group field theories for all loop quantum gravity}",
    eprint = "1409.3150",
    archivePrefix = "arXiv",
    primaryClass = "gr-qc",
    reportNumber = "AEI-2014-044",
    doi = "10.1088/1367-2630/17/2/023042",
    journal = "New J. Phys.",
    volume = "17",
    number = "2",
    pages = "023042",
    year = "2015"
}

@article{Finocchiaro:2020fhl,
    author = "Finocchiaro, Marco and Oriti, Daniele",
    title = "{Renormalization of Group Field Theories for Quantum Gravity: New Computations and Some Suggestions}",
    eprint = "2004.07361",
    archivePrefix = "arXiv",
    primaryClass = "hep-th",
    doi = "10.3389/fphy.2020.552354",
    journal = "Front. in Phys.",
    volume = "8",
    pages = "552354",
    year = "2021"
}

@article{millington,
  title = {Symmetries and conservation laws in non-Hermitian field theories},
  author = {Alexandre, Jean and Millington, Peter and Seynaeve, Dries},
  journal = {Phys. Rev. D},
  volume = {96},
  issue = {6},
  pages = {065027},
  numpages = {9},
  year = {2017},
  month = {Sep},
  publisher = {American Physical Society},
  doi = {10.1103/PhysRevD.96.065027},
  url = {https://link.aps.org/doi/10.1103/PhysRevD.96.065027}
}

@article{millington2,
doi = {10.1088/1742-6596/952/1/012012},
url = {https://doi.org/10.1088/1742-6596/952/1/012012},
year = {2018},
month = {jan},
publisher = {IOP Publishing},
volume = {952},
number = {1},
pages = {012012},
author = {Alexandre, Jean and Millington, Peter and Seynaeve, Dries},
title = {Consistent description of field theories with non-Hermitian mass terms},
journal = {Journal of Physics: Conference Series}
}

@article{Ashida:2020dkc,
    author = "Ashida, Yuto and Gong, Zongping and Ueda, Masahito",
    title = "{Non-Hermitian physics}",
    eprint = "2006.01837",
    archivePrefix = "arXiv",
    primaryClass = "cond-mat.mes-hall",
    doi = "10.1080/00018732.2021.1876991",
    journal = "Adv. Phys.",
    volume = "69",
    number = "3",
    pages = "249--435",
    year = "2021"
}

@phdthesis{Seynaeve:2020oza,
    author = "Seynaeve, Dries",
    title = "{Non-Hermitian Quantum Field Theory}",
    school = "King's Coll. London",
    year = "2020"
}

@book{kloeden,
author = {Kloeden, Peter and Rasmussen, Martin},
year = {2011},
month = {08},
pages = {},
title = {Nonautonomous Dynamical Systems},
volume = {176},
isbn = {978-0-8218-6871-3},
journal = {Math. Surveys Monogr.},
doi = {10.1090/surv/176}
}

@book{chepyzhov2002attractors,
  title={Attractors for Equations of Mathematical Physics},
  author={Chepyzhov, V.V. and Vishik, M.I.},
  isbn={9780821829509},
  lccn={01046406},
  series={American Mathematical Society Colloquium publications},
  url={https://books.google.co.jp/books?id=qCRhhHQGfrcC},
  year={2002},
  publisher={American Mathematical Society}
}

@article{Vishik_2011,
doi = {10.1070/RM2011v066n04ABEH004753},
url = {https://doi.org/10.1070/RM2011v066n04ABEH004753},
year = {2011},
month = {aug},
publisher = {},
volume = {66},
number = {4},
pages = {637},
author = {Vishik, Marko I and Chepyzhov, Vladimir V},
title = {Trajectory attractors of equations of mathematical physics},
journal = {Russian Mathematical Surveys}
}

@article{Loll:2019rdj,
    author = "Loll, R.",
    title = "{Quantum Gravity from Causal Dynamical Triangulations: A Review}",
    eprint = "1905.08669",
    archivePrefix = "arXiv",
    primaryClass = "hep-th",
    doi = "10.1088/1361-6382/ab57c7",
    journal = "Class. Quant. Grav.",
    volume = "37",
    number = "1",
    pages = "013002",
    year = "2020"
}

@article{Ambjorn:2012jv,
    author = "Ambjorn, J. and Goerlich, A. and Jurkiewicz, J. and Loll, R.",
    title = "{Nonperturbative Quantum Gravity}",
    eprint = "1203.3591",
    archivePrefix = "arXiv",
    primaryClass = "hep-th",
    doi = "10.1016/j.physrep.2012.03.007",
    journal = "Phys. Rept.",
    volume = "519",
    pages = "127--210",
    year = "2012"
}

@article{Baratin:2011hp,
    author = "Baratin, Aristide and Oriti, Daniele",
    title = "{Group field theory and simplicial gravity path integrals: A model for Holst-Plebanski gravity}",
    eprint = "1111.5842",
    archivePrefix = "arXiv",
    primaryClass = "hep-th",
    reportNumber = "LPT-ORSAY-11-120",
    doi = "10.1103/PhysRevD.85.044003",
    journal = "Phys. Rev. D",
    volume = "85",
    pages = "044003",
    year = "2012"
}

@article{Marchetti:2020umh,
    author = "Marchetti, Luca and Oriti, Daniele",
    title = "{Effective relational cosmological dynamics from Quantum Gravity}",
    eprint = "2008.02774",
    archivePrefix = "arXiv",
    primaryClass = "gr-qc",
    doi = "10.1007/JHEP05(2021)025",
    journal = "JHEP",
    volume = "05",
    pages = "025",
    year = "2021"
}

@article{Marchetti:2021gcv,
    author = "Marchetti, Luca and Oriti, Daniele",
    title = "{Effective dynamics of scalar cosmological perturbations from quantum gravity}",
    eprint = "2112.12677",
    archivePrefix = "arXiv",
    primaryClass = "gr-qc",
    doi = "10.1088/1475-7516/2022/07/004",
    journal = "JCAP",
    volume = "07",
    number = "07",
    pages = "004",
    year = "2022"
}

@article{Marchetti:2020qsq,
    author = "Marchetti, Luca and Oriti, Daniele",
    title = "{Quantum fluctuations in the effective relational GFT cosmology}",
    eprint = "2010.09700",
    archivePrefix = "arXiv",
    primaryClass = "gr-qc",
    doi = "10.3389/fspas.2021.683649",
    journal = "Front. Astron. Space Sci.",
    volume = "8",
    pages = "683649",
    year = "2021"
}

@article{Calcinari:2023sax,
    author = "Calcinari, Andrea and Gielen, Steffen",
    title = "{Generalized Gaussian states in group field theory and su(1,1) quantum cosmology}",
    eprint = "2310.08667",
    archivePrefix = "arXiv",
    primaryClass = "gr-qc",
    doi = "10.1103/PhysRevD.109.066022",
    journal = "Phys. Rev. D",
    volume = "109",
    number = "6",
    pages = "066022",
    year = "2024"
}

@article{Ladstatter:2025kgu,
    author = {Ladst{\"a}tter, Tom R. and Marchetti, Luca},
    title = "{Interacting Scalar Field Cosmology from Full Quantum Gravity}",
    eprint = "2508.16194",
    archivePrefix = "arXiv",
    primaryClass = "gr-qc",
    month = "8",
    year = "2025"
}

@article{Calcinari:2024pek,
    author = "Calcinari, Andrea and Gielen, Steffen",
    title = "{Relational dynamics and Page-Wootters formalism in group field theory}",
    eprint = "2407.03432",
    archivePrefix = "arXiv",
    primaryClass = "gr-qc",
    doi = "10.22331/q-2025-01-27-1610",
    journal = "Quantum",
    volume = "9",
    pages = "1610",
    year = "2025"
}

@article{Mukhanov:2013tua,
    author = "Mukhanov, Viatcheslav",
    title = "{Quantum Cosmological Perturbations: Predictions and Observations}",
    eprint = "1303.3925",
    archivePrefix = "arXiv",
    primaryClass = "astro-ph.CO",
    doi = "10.1140/epjc/s10052-013-2486-7",
    journal = "Eur. Phys. J. C",
    volume = "73",
    pages = "2486",
    year = "2013"
}

@article{Carrozza:2016vsq,
    author = "Carrozza, Sylvain",
    title = "{Flowing in Group Field Theory Space: a Review}",
    eprint = "1603.01902",
    archivePrefix = "arXiv",
    primaryClass = "gr-qc",
    doi = "10.3842/SIGMA.2016.070",
    journal = "SIGMA",
    volume = "12",
    pages = "070",
    year = "2016"
}

@article{Pithis:2020kio,
    author = {Pithis, Andreas G. A. and Th\"urigen, Johannes},
    title = "{Phase transitions in TGFT: functional renormalization group in the cyclic-melonic potential approximation and equivalence to O$(N)$ models}",
    eprint = "2009.13588",
    archivePrefix = "arXiv",
    primaryClass = "hep-th",
    doi = "10.1007/JHEP12(2020)159",
    journal = "JHEP",
    volume = "12",
    pages = "159",
    year = "2020"
}

@article{Jercher:2021bie,
    author = "Jercher, Alexander F. and Oriti, Daniele and Pithis, Andreas G. A.",
    title = "{Emergent cosmology from quantum gravity in the Lorentzian Barrett-Crane tensorial group field theory model}",
    eprint = "2112.00091",
    archivePrefix = "arXiv",
    primaryClass = "gr-qc",
    doi = "10.1088/1475-7516/2022/01/050",
    journal = "JCAP",
    volume = "01",
    number = "01",
    pages = "050",
    year = "2022"
}

@phdthesis{Jordan:2013sok,
    author = "Jordan, Samo",
    title = "{Globally and locally causal dynamical triangulations}",
    school = "Nijmegen U., Nijmegen U.",
    year = "2013"
}

@article{Gielen:2019kae,
    author = "Gielen, Steffen and Polaczek, Axel",
    title = "{Generalised effective cosmology from group field theory}",
    eprint = "1912.06143",
    archivePrefix = "arXiv",
    primaryClass = "gr-qc",
    doi = "10.1088/1361-6382/ab8f67",
    journal = "Class. Quant. Grav.",
    volume = "37",
    number = "16",
    pages = "165004",
    year = "2020"
}

@article{Loll:1998aj,
    author = "Loll, Renate",
    title = "{Discrete approaches to quantum gravity in four-dimensions}",
    eprint = "gr-qc/9805049",
    archivePrefix = "arXiv",
    reportNumber = "AEI-063A, AEI-063",
    doi = "10.12942/lrr-1998-13",
    journal = "Living Rev. Rel.",
    volume = "1",
    pages = "13",
    year = "1998"
}

@article{Gielen:2016uft,
    author = "Gielen, Steffen",
    title = "{Emergence of a low spin phase in group field theory condensates}",
    eprint = "1604.06023",
    archivePrefix = "arXiv",
    primaryClass = "gr-qc",
    reportNumber = "IMPERIAL-TP-2016-SG-2",
    doi = "10.1088/0264-9381/33/22/224002",
    journal = "Class. Quant. Grav.",
    volume = "33",
    number = "22",
    pages = "224002",
    year = "2016"
}

@article{Gielen:2016dss,
    author = "Gielen, Steffen and Sindoni, Lorenzo",
    title = "{Quantum Cosmology from Group Field Theory Condensates: a Review}",
    eprint = "1602.08104",
    archivePrefix = "arXiv",
    primaryClass = "gr-qc",
    reportNumber = "IMPERIAL-TP-2016-SG-1",
    doi = "10.3842/SIGMA.2016.082",
    journal = "SIGMA",
    volume = "12",
    pages = "082",
    year = "2016"
}

@article{Marchetti:2022igl,
    author = {Marchetti, Luca and Oriti, Daniele and Pithis, Andreas G. A. and Th\"urigen, Johannes},
    title = "{Phase transitions in TGFT: a Landau-Ginzburg analysis of Lorentzian quantum geometric models}",
    eprint = "2209.04297",
    archivePrefix = "arXiv",
    primaryClass = "gr-qc",
    doi = "10.1007/JHEP02(2023)074",
    journal = "JHEP",
    volume = "02",
    pages = "074",
    year = "2023"
}

@article{Jercher:2023nxa,
    author = "Jercher, Alexander F. and Marchetti, Luca and Pithis, Andreas G. A.",
    title = "{Scalar cosmological perturbations from quantum entanglement within Lorentzian quantum gravity}",
    eprint = "2308.13261",
    archivePrefix = "arXiv",
    primaryClass = "gr-qc",
    doi = "10.1103/PhysRevD.109.066021",
    journal = "Phys. Rev. D",
    volume = "109",
    number = "6",
    pages = "066021",
    year = "2024"
}

@article{Jercher:2023kfr,
    author = "Jercher, Alexander F. and Marchetti, Luca and Pithis, Andreas G. A.",
    title = "{Scalar Cosmological Perturbations from Quantum Gravitational Entanglement}",
    eprint = "2310.17549",
    archivePrefix = "arXiv",
    primaryClass = "gr-qc",
    month = "10",
    year = "2023"
}

@inproceedings{Oriti:2021oux,
    author = "Oriti, Daniele",
    title = "{Tensorial Group Field Theory condensate cosmology as an example of spacetime emergence in quantum gravity}",
    eprint = "2112.02585",
    archivePrefix = "arXiv",
    primaryClass = "gr-qc",
    month = "12",
    year = "2021"
}

@article{Marchetti:2024tjq,
    author = "Marchetti, Luca and Mehmood, Hassan and Husain, Viqar",
    title = "{Exactly soluble group field theory}",
    eprint = "2412.09851",
    archivePrefix = "arXiv",
    primaryClass = "gr-qc",
    doi = "10.1103/v1rr-bhqx",
    journal = "Phys. Rev. D",
    volume = "112",
    number = "4",
    pages = "044056",
    year = "2025"
}

@article{Marchetti:2024nnk,
    author = "Marchetti, Luca and Wilson-Ewing, Edward",
    title = "{Relational Observables in Group Field Theory}",
    eprint = "2412.14622",
    archivePrefix = "arXiv",
    primaryClass = "gr-qc",
    month = "12",
    year = "2024"
}

@article{Gielen:2024sxs,
    author = "Gielen, Steffen",
    title = "{Hilbert space formalisms for group field theory}",
    eprint = "2412.07847",
    archivePrefix = "arXiv",
    primaryClass = "gr-qc",
    doi = "10.1088/1361-6382/adc655",
    journal = "Class. Quant. Grav.",
    volume = "42",
    number = "8",
    pages = "083001",
    year = "2025"
}
\end{document}